\documentclass[preprints,article,accept,moreauthors,color,dvi2pdf]{mdpi}
\firstpage{1}
\pubvolume{xx}
\issuenum{1}
\articlenumber{5}
\pubyear{2022}
\copyrightyear{2022}
\usepackage{amssymb}
\usepackage{amsmath}
\usepackage{amsfonts}
\usepackage{epsfig}
\usepackage{graphicx}
\usepackage{tabularx}
\usepackage{verbatim}
\usepackage{xcolor}
\newcommand{\eq}{\begin{eqnarray}}
\newcommand{\en}{\end{eqnarray}}

\history{Received: 30 March 2022; Accepted: 27 June 2022; Published: 30 June 2022}

\continuouspages{yes}

\Title{Coupled-channel analysis of the process $\gamma \gamma \to \pi^0 \pi^0$}

\Author{Yury~S.~Surovtsev$^{1}$,
        Petr~Byd\v{z}ovsk\'y$^{2}$,
        Thomas~Gutsche$^{3}$,
        Robert~Kami\'nski$^{4}$,
        Valery~E.~Lyubovitskij$^{3,5,6,7,8}$,
        Miroslav~Nagy$^{9}$
}

\address{
$^{1}$ \quad
Bogoliubov Laboratory of Theoretical Physics,
Joint Institute for Nuclear Research, 141980 Dubna, Russia \\
$^{2}$ \quad Nuclear Physics Institute of the AS CR, 25068 \v{R}e\v{z},
Czech Republic \\
$^{3}$ \quad
Institut f\"ur Theoretische Physik,
Universit\"at T\"ubingen,
Kepler Center for Astro and Particle Physics,
Auf der Morgenstelle 14, D-72076 T\"ubingen, Germany \\
$^{4}$ \quad
Institute of Nuclear Physics PAS, Cracow 31342, Poland \\
$^{5}$ \quad
Departamento de F\'\i sica y Centro Cient\'\i fico
Tecnol\'ogico de Valpara\'\i so-CCTVal, Universidad T\'ecnica
Federico Santa Mar\'\i a, Casilla 110-V, Valpara\'\i so, Chile \\
$^{6}$ \quad
Millennium Institute for Subatomic Physics at
the High-Energy Frontier (SAPHIR) of ANID,
Fern\'andez Concha 700, Santiago, Chile \\
$^{7}$ \quad
Department of Physics, Tomsk State University,
634050 Tomsk, Russia \\
$^{8}$ \quad
Tomsk Polytechnic University, 634050 Tomsk, Russia \\
$^{9}$ \quad
Institute of Physics, SAS, Bratislava 84511, Slovak Republic}

\corres{Correspondence: valeri.lyubovitskij@uni-tuebingen.de}

\abstract{
We study the process $\gamma\gamma\to\pi^0\pi^0$ involving the principal mechanisms,
the structure of its cross section and the role of individual isoscalar-tensor resonances
in the saturation of its energy spectrum.}


\begin{document}

\section{Introduction}

The observation of the processes $\gamma\gamma\to hadrons$ (including $\gamma\gamma\to\pi^0\pi^0$ by
the Crystal Ball~\cite{CrBall_gg} and Belle~\cite{Belle13,Belle_gg}) is rather interesting since they
involve an immediate transformation of energy to the masses of strongly interacting particles.
The importance of this phenomenon is obvious. Therefore, when analyzing data on energy spectrum
of cross section of process $\gamma\gamma\to\pi^0\pi^0$ in addition to determination of
dipole and quadrupole polarizabilities of pions (see, e.g., Ref.~\cite{KBP}), it is extremely
interesting to study the coupled-channels contributions, including interference phenomena,
and also the role of individual resonances in saturation of the energy spectrum of this process.

Generally processes $\gamma\gamma\to\pi^+\pi^-,\pi^0\pi^0$ were already studied earlier
(see, e.g., Refs.~\cite{Garcia}-\cite{Lu} and references therein), especially after appearance
of the experimental data on these reactions \cite{CrBall_gg,Belle13,Belle_gg}. In these works one
considers productions of pion pairs in annihilation of both real and virtual photons in the dispersion
relation approach with imposing various constraints, such as a Mandelstam analyticity and chiral constraints.

For the observed two-photon annihilation the $\pi^0\pi^0$ pairs are produced in the $S$- and $D$-wave states
with isospins $I=0$ and $2$. The isoscalar parts dominate significantly. Earlier we have
shown~\cite{Surovtsev:2012vg} that physical resonances, as the scalar and tensor ones, can be considered
correctly only in approaches where the $S$-matrix is determined on the multi-sheeted Riemann surface
(the isoscalar-scalar resonances on an 8-sheeted surface, isoscalar-tensor ones on a 16-sheeted one).
One should keep in mind that the dispersion relations are written on the 2-sheeted Riemann
surface.
Therefore, we consider this process by allowing for coupled channels.
For the process $\gamma\gamma\to\pi\pi$ with rescattering in the final state the coupled channels are
$\gamma\gamma\to(\pi\pi,K\overline{K},\eta\eta)\to\pi\pi$. The amplitudes for the isoscalar $S$-wave
three-channel $\pi\pi$ scattering are taken from our model-independent analysis of data for
$\pi\pi\to\pi\pi,K\overline{K},\eta\eta$~\cite{SBLKN-prd14}.
Our approach is based on analyticity and
unitarity by using a uniformization procedure (for detailed review of this method
see, e.g., Refs.~\cite{Krupa:1984qr,KMS-96,SBGLKN-HQ2013}). For physical applications see
Refs.~\cite{Surovtsev:2012vg}-\cite{SBL-prd12}.
This method allows to study
the $S$-matrix elements in multichannel hadron scattering processes with taking into account
a structure of the corresponding Riemann surface. In particular, in the two- and three-channel
cases the corresponding $S$-matrices are determined on the 4- and 8-sheeted Riemann surfaces,
respectively, which are transformed onto the uniformization plane by corresponding conformal
mappings. Then we have a possibility to represent correctly multichannel resonances by poles
(and by corresponding zeros) on all sheets of the Riemann surface.

Our previous combined description of data on the decays of charmonia --
$J/\psi\to\phi(\pi\pi,K\overline{K})$, $\psi(2S)\to J/\psi\,\pi\pi$, and $X(4260)\to J/\psi~\pi^+\pi^-$
-- and of bottomonia -- $\Upsilon(mS)\to\Upsilon(nS)\pi\pi$
($m>n$, $m=2,3,4,5,$ $n=1,2,3$)~\cite{SBGKLN-prd18}
(where the contributing amplitude of isoscalar $S$-wave three-channel $\pi\pi$ scattering was directly
taken from the analysis of the processes $\pi\pi\to\pi\pi,K\overline{K},\eta\eta$~\cite{SBLKN-prd14}) --
implied a confirmation of all our earlier results on the scalar
mesons. Therefore, when considering the cross section of the process $\gamma\gamma\to\pi^0\pi^0$, we will
keep the parameters of the $S$-wave three-channel $\pi\pi$ scattering amplitude~\cite{SBLKN-prd14}
unchanged, hoping to obtain an additional confirmation of our results on the scalar mesons.

For the $D$-wave of multi-channel $\pi\pi$-scattering in $\gamma\gamma\to\pi\pi$ we will use the results
of our 4-channel analysis of the processes $\pi\pi\to\pi\pi,(2\pi)(2\pi),K\overline{K},\eta\eta$, where
we also considered  the $(2\pi)(2\pi)$ channel~\cite{SBKN-prd10} explicitly. As in the case of the
$S$-wave 3-channel $\pi\pi$-scattering, here the resonances are represented by poles on the
16-sheeted Riemann surface. The resonance poles are generated by some 4-channel Breit--Wigner forms
with a Blatt--Weisskopf barrier factor due to the meson spins.

In the PDG issue 2020~\cite{PDG} twelve resonances in the isoscalar-tensor meson sector
are discussed, $f_2(1270)$, $f_2(1430)^*$, $f_2^\prime(1525)$, $f_2(1565)^*$,
$f_2(1640)^*$, $f_2(1810)^*$, $f_2(1910)^*$, $f_2(1950)$, $f_2(2010)$, $f_2(2150)^*$,
$f_2(2300)$, and $f_2(2340)$. The resonances denoted with asterisk were omitted
from the summary table as they still need an experimental confirmation. There is considered also
the state $f_J(2220)$ ($J^{PC}=2^{++}~{\rm or}~4^{++}$) which was omitted from the summary table.
In our previous analysis of the multichannel $\pi\pi$ scattering with eleven states,
performed in 2010~\cite{SBKN-prd10}, we have not considered the states $f_2(1640)$, $f_2(1910)$,
$f_2(2150)$, $f_2(2300)$, and $f_2(2340)$ but have included the states $f_2(1730)$,
$f_2(2020)$, $f_2(2240)$, and  $f_2(2410)$.
Considering these eleven resonances we obtained a satisfactory combined description of data on the
D-wave processes $\pi\pi\to\pi\pi,K\overline{K},\eta\eta$, explicitly allowing for the
$(2\pi)(2\pi)$ channel.
The state $f_2(2020)$ practically did not change the description of data but it
allowed the interpretation of the $f_2(2000)$ state as a glueball as it was done
in Ref.~\cite{Ani05} using an approach based on the $1/N_c$-expansion.
In the presented analysis we use these eleven
resonances for the isoscalar-tensor meson sector.

The paper is organized as follows.
In Sec.~II we derive the formalism for taking into account
effects of multi-channel $\pi\pi$ scattering in the decay mode
$\gamma\gamma \to \pi^0\pi^0$. First we present some general basic formulas and a model for the process
$\gamma\gamma\to\pi^0\pi^0$, which is a development of the one proposed in Refs.~\cite{Au,Boglione},
however, with allowing for our previous results on the multichannel $\pi\pi$ scattering,
that for obtaining correct values for the  $f_0$-resonance parameters in the analysis of
multichannel $\pi\pi$ scattering data it is needed, as minimum, the tree-coupled channel analysis,
namely the combined analysis of the data on $S$-wave processes
$\pi\pi\to\pi\pi,K\overline{K},\eta\eta$~\cite{SBLKN-jpgnpp14}.
Analogously, to obtain the  $f_2$-resonance parameters,
it is needed the four-channel analysis of the data on $D$-wave processes
$\pi\pi\to\pi\pi,(2\pi)(2\pi),K\overline{K},\eta\eta$~\cite{SBKN-prd10}.
Further we outline our model-independent $S$-matrix approach based on first principles,
such as analyticity and unitarity~\cite{SBKN-prd10}. It was used in calculating
the $S$- and $D$-wave amplitudes of the above-indicated coupled processes,
which are applied in the model for the process $\gamma\gamma\to\pi^0\pi^0$.
Note that for the present it is reasonable not to consider the channel with isospin $I=2$
because the Born approximation is equal to zero and there are no mesonic resonances
with $I=2$. Taking into account the format of this paper, we avoid the excessive details
of formalism, dispatching to the corresponding references.
In Sec.~III we show the results of calculations of the cross section energy spectrum of
the $\gamma\gamma\to\pi^0\pi^0$ in the presented model when comparing with the experimental
data~\cite{CrBall_gg,Belle13,Belle_gg}. There are investigated separately contributions
of the $S$- and $D$-waves, of the individual channels to the energy spectrum. Also we
show the calculated energy spectra of the $\gamma\gamma\to\pi^0\pi^0$ when switching off
the individual $f_2$ resonances, grouped around the energy interval $1.5-1.73$~GeV,
where our calculations diverge with the data. Since our approach of principle consists
in the combined description of this process and of the above-indicated processes of
the $S$- and $D$-wave multichannel $\pi\pi$ scattering, we carry out and show the results
of the combined analysis (for each case of switching off the individual $f_2$ resonances)
of $\gamma\gamma\to\pi^0\pi^0$, of the isoscalar $S$-wave processes
$\pi\pi\to\pi\pi,K\overline{K},\eta\eta$ and of the isoscalar $D$-wave processes
$\pi\pi\to\pi\pi,(2\pi)(2\pi),K\overline{K},\eta\eta$.
Finally, Sec.~IV contains our conclusions.

\section{Effect of multi-channel $\pi\pi$ scattering in $\gamma \gamma \to \pi^0 \pi^0$}

When studying the process $\gamma\gamma\to\pi^0\pi^0$ we have to include the effect of intermediate states.
Since photons couple to charged objects production of $\pi^0\pi^0$ pairs can only occur through the formation
of $\pi^+\pi^-$ and $K^+K^-$ pairs in the intermediate states of the final-state rescattering processes.
We describe the coupling of photon with charged pseudoscalar mesons based on the vector meson dominance
(VMD) model. According to the VMD model, the hadron electromagnetic
current is given by the linear combination of individual contributions of vector meson fields \cite{Sakurai}:
\begin{equation}\label{MVD}
J_\mu(x)=-\left[\frac{m_\rho^2}{2\gamma_\rho}\rho^0_\mu(x)+\frac{m_\omega^2}{2\gamma_\omega}\omega^0_\mu(x)
+\frac{m_\phi^2}{2\gamma_\phi}\phi^0_\mu(x)\right],
\end{equation}
where $m_V$ is the mass of vector mesons ($V=\rho^0,\omega,\phi$). The coupling constants of the photon with
vector mesons are determined by the normalization conditions of vector fields:
$$<0|J_\mu(0)|V>=-\frac{m_V^2}{2\gamma_V}V_\mu.$$
In the $SU_3$-symmetry scheme with ideal $\omega-\phi$ meson mixing one has the relations:
\begin{equation}\label{MVDrelation}
\gamma_\rho:\gamma_\omega:\gamma_\phi=1:3:\left(-\frac{3}{\sqrt{2}}\right).
\end{equation}
For the constants $\gamma_V$ there are found the following values \cite{Schildknecht73}:
\begin{equation}\label{MVD2}
\frac{\gamma_\rho^2}{4\pi}=0.64\pm0.05\,,\quad
\frac{\gamma_\omega^2}{4\pi}=4.8\pm0.5,\quad \frac{\gamma_\phi^2}{4\pi}=2.8\pm0.2.
\end{equation}
The use of the VMD model permits us to understand how the intermediate states
are formed in the process $\gamma\gamma\to\pi^0\pi^0$. According to the VMD hypothesis,
one can consider the $\gamma$-quant state as a superposition of
the $\rho^0$-, $\omega$- and $\phi$-meson states~\cite{AkhRek}:
\begin{equation}\label{MVDstates}
\left|\gamma\right>=\sqrt{\alpha\pi}\left(\gamma_\rho^{-1}\left|\rho^0\right>+
\gamma_\omega^{-1}\left|\omega\right>-\gamma_\phi^{-1}\left|\phi\right>\right).
\end{equation}
It is seen that in the $\gamma \gamma$ annihilation the $\rho^0 \rho^0$ components
are responsible for production of two pairs of charged pions $(\pi^+ \pi^-)$,
the $\phi$$\phi$ for $(K^+ K^-)$, and $\omega$$\omega$ for two triplets $(\pi^+ \pi^- \pi^0)$.
At that, it is assumed that two opposite-charge mesons (each from the different
pairs/triplets of particles, produced by different photons) are paired into $t$-channel propagator,
giving the nearest mesons, exchanged in the crossing-channels. Then in the intermediate
states we have two/four pseudoscalar mesons. The VMD relations [see Eqs.~(\ref{MVDrelation})
and~(\ref{MVD2})] are used to define the coupling which is attached to the corresponding diagrams
describing the processes $\gamma + \gamma \to \rho^0 + \rho^0 \to \pi^+ + \pi^-$, 
$\gamma + \gamma \to \phi + \phi \to K^+ + K^-$, 
$\gamma + \gamma \to \omega + \omega \to 2 \pi^+ + 2 \pi^-$, and  
$\gamma + \gamma \to \omega + \omega \to \pi^+ + \pi^- + 2 \pi^0$. 

The cross section of process $\gamma\gamma\to\pi^0\pi^0$ is written via the helicity amplitudes $M_{++}$
and $M_{+-}$ (the subscripts label the helicities of the incoming photons)~\cite{Abarbanel,Morgan-Penn}
as follows
\begin{equation}
\frac{d\sigma}{d\Omega}=\frac{1}{256\pi^2s}\sqrt{\frac{s-4m_\pi^2}{s}}\left(|M_{++}|^2+|M_{+-}|^2\right),
\end{equation}
where $s=(k_1+k_2)^2$ with $k_1$ and $k_2$ being the 4-momenta of the photons.
These helicity amplitudes are decomposed into partial waves:
\begin{eqnarray}\label{hel-ampl}
&&M_{++}(s,\theta,\phi)=e^2\sqrt{16\pi}\sum\limits_{J\geq 0}F_{J0}(s)Y_{J0}(\theta,\phi),\\
&&M_{+-}(s,\theta,\phi)=e^2\sqrt{16\pi}\sum\limits_{J\geq 2}F_{J2}(s)Y_{J2}(\theta,\phi).
\end{eqnarray}
The partial wave amplitudes $F_{J\lambda}$~(helicity $\lambda=0,2$) must be determined in the analysis.
In the following we do the truncation of the partial wave expansion in Eq.~(\ref{hel-ampl}) by
including only the first leading term following the idea proposed in Ref.\cite{Boglione}. This is related 
to the fact, that in the final state $\pi\pi$ interaction, 
contributing to the process $\gamma\gamma\to\pi^0\pi^0$, the S- and D-wave dominate.

In Refs.~\cite{Au,Boglione}
it is argued that the partial amplitudes are approximated in the following form:
\begin{equation}\label{eq:1}
F_{J\lambda}^{I=0}(\gamma\gamma\to\pi\pi)=\sum\limits_{n}
a_{J\lambda}^{I=0}(\gamma\gamma\to n)T_{J}^{I=0}(n\to\pi\pi),
\end{equation}
where the $a_{J\lambda}^{I=0}(\gamma\gamma\to n)$ approximate transitions from $\gamma\gamma$
to the intermediate states; they are functions of $s$ and real above the $\pi\pi$ threshold.
Note that it is reasonable in the case of process $\gamma\gamma\to\pi^0\pi^0$ not to consider
the channel with isospin $I=2$ because the Born approximation is equal to zero and there are
no mesonic resonances with $I=2$. The $T_{J}^I(n\to\pi\pi)$, describing transitions from
the intermediate states into pion pair, satisfy the unitarity conditions:
\begin{equation}\label{T:unitarity}
{\rm Im}T_{J}^I(n\to\pi\pi)=\sum\limits_{n^\prime}\rho_nT_{J}^I(n\to n^\prime)^*T_{J}^I(n^\prime\to\pi\pi),
\end{equation}
where the sum is over the hadronic intermediate states $n^\prime$, kinematically admitted; $\rho_n$
is the corresponding phase space factor for each channel. When using this relation, one can see,
that the expression~(\ref{eq:1}) satisfies the unitarity condition for the amplitude
$F_{J\lambda}^{I}(\gamma\gamma\to\pi\pi)$~\cite{Au,Boglione}:
\begin{equation}\label{F:unitarity}
{\rm Im}F_{J\lambda}^{I}(\gamma\gamma\to\pi\pi)=\sum\limits_{n}\rho_n F_{J\lambda}^{I}(\gamma\gamma
\to n)^*T_{J}^{I}(n\to\pi\pi),
\end{equation}
Considering $S$-wave multichannel $\pi\pi$ scattering in the final $\pi^0\pi^0$ state, we shall deal
with the 3-channel case, i.e. with the reactions $\pi\pi\to\pi\pi,K\overline{K},\eta\eta$.
In Ref.~\cite{SBLKN-jpgnpp14} it was shown that this is a minimal number of coupled channels needed
to obtain correct values for the $f_0$-resonance parameters in the analysis of multichannel $\pi\pi$
scattering data. For the $D$-wave multichannel $\pi\pi$ scattering one ought to consider, as minimum,
four coupled channels:
$\pi\pi\to\pi\pi,(2\pi)(2\pi),K\overline{K},\eta\eta$~\cite{SBKN-prd10}.
We use the following relation between the $T$-matrix and $S$-matrix
$(T^0_0)_{ij} = \sqrt{s}/(4ik) \, [(S^0_0)_{ij} - 1]$.
Therefore, redenoting in Eq.~(\ref{eq:1}) $a_{J=0\lambda=0}^{I=0}(\gamma\gamma\to n)\equiv
a(\gamma\gamma\to n)$ and $a_{J=2\lambda=2}^{I=0}(\gamma\gamma\to n)\equiv b(\gamma\gamma\to n)$,
we write the $S$- and $D$-wave amplitudes as
\begin{eqnarray}\label{S-D-ampl}
F_{00}^{I=0}(\gamma\gamma\to\pi^0\pi^0)
&=&a(\gamma\gamma\to\pi^+\pi^-)T_{0}^{I=0}(\pi^+\pi^-\to\pi^0\pi^0)\nonumber\\
&+&a(\gamma\gamma\to K^+K^-)T_{0}^{I=0}(K^+K^-\to\pi^0\pi^0)\nonumber\\
&+&a(\gamma\gamma\to\eta\eta)T_{0}^{I=0}(\eta\eta\to\pi^0\pi^0)\,,\\
F_{22}^{I=0}(\gamma\gamma\to\pi^0\pi^0)
&=&b(\gamma\gamma\to\pi^+\pi^-)T_{2}^{I=0}(\pi^+\pi^-\to\pi^0\pi^0)\nonumber\\
&+&b(\gamma\gamma\to2\pi^+2\pi^-)T_{2}^{I=0}(2\pi^+2\pi^-\to\pi^0\pi^0)\nonumber\\
&+&b(\gamma\gamma\to K^+K^-)T_{2}^{I=0}(K^+K^-\to\pi^0\pi^0)\nonumber\\
&+&b(\gamma\gamma\to\eta\eta)T_{2}^{I=0}(\eta\eta\to\pi^0\pi^0)\,,
\label{S-D-ampl2}
\end{eqnarray}
where
\begin{eqnarray}\label{contrib}
&&a(\gamma\gamma\to\pi^+\pi^-)=\frac{\alpha_{10}}{(s-\gamma_1)^2}+\alpha_{11}+\alpha_{12}s,\nonumber\\
&&a(\gamma\gamma\to K^+K^-)=\frac{\beta_{20}}{(s-\gamma_2)^2}+\beta_{21}+\beta_{22}s,\nonumber\\
&&a(\gamma\gamma\to\eta\eta)=\beta_{31}+\beta_{32}s,\nonumber\\
&&b(\gamma\gamma\to\pi^+\pi^-)=\frac{\delta_{10}}{(s-\rho_1)^2}+\delta_{11}+\delta_{12}s,\\
&&b(\gamma\gamma\to2\pi^+2\pi^-)=\frac{\delta_{20}}{(s-\rho_2)^2}+\delta_{21}+\delta_{22}s,\nonumber\\
&&b(\gamma\gamma\to K^+K^-)=\delta_{31}+\delta_{32}s,\nonumber\\
&&b(\gamma\gamma\to\eta\eta)=\delta_{41}+\delta_{42}s\;\nonumber
\end{eqnarray}
with an obvious notation. 
These parameters must be determined, on the whole, in a combined fit
to data on $\gamma\gamma\to\pi^0\pi^0$~\cite{CrBall_gg,Belle13,Belle_gg} and on
isoscalar $S$- and $D$-wave multi-channel $\pi\pi$-scattering. 
Note that the VMD couplings are absorbed into the parameters
$\alpha_{11}$, $\alpha_{12}$, $\beta_{21}$, $\beta_{22}$, 
$\delta_{ij}$, in Eq.~(\ref{contrib}) and the latter 
are fitted to data to guarantee that the obtained corresponding values of these couplings 
approximately satisfy to the relations (\ref{MVDrelation}) and (\ref{MVD2}). 
Let us explain the pole terms in
the coefficients $a$ and $b$. Whereas in the direct channel of process $\gamma\gamma\to\pi\pi$ the
isoscalar-scalar and isoscalar-tensor meson resonances contribute, in the crossing-channels
(the Compton scattering $\gamma\pi\to\gamma\pi$) other resonances contribute: from the nearest
-- exchanges of the $\rho(770)$, $\omega(782)$, $\phi(1020)$, $b_1(1235)$, $a_1(1260)$ and $a_2(1320)$.
We approximate these exchanges effectively by the pole term determined in the analysis.
Also the pole term in $a(\gamma\gamma\to K^+K^-)$ approximates the $K$-meson exchange in
the $K^+K^-$ intermediate state between the $\gamma K^+$ and $\gamma K^-$ vertices; the pole terms
in $b(\gamma\gamma\to\pi^+\pi^-)$ and in $b(\gamma\gamma\to2\pi^+2\pi^-)$ arise from pion exchanges
in the $\pi^+\pi^-$ and $4\pi$ intermediate states, respectively, between the $\gamma\pi^+$ and
$\gamma\pi^-$ vertices and between the $\gamma2\pi$ vertices.
Note that the appearance of the 2nd-order poles is related
with the fact that both channels, being crossing with respect to
the direct channels, coincide with each other.

We further complement our previous satisfactory combined description of data on the decays of
charmonia -- $J/\psi\to\phi(\pi\pi,K\overline{K})$, $\psi(2S)\to J/\psi\,\pi\pi$ and $X(4260)\to
J/\psi~\pi^+\pi^-$, of bottomonia -- $\Upsilon(mS)\to\Upsilon(nS)\pi\pi$ ($m>n$, $m=2,3,4,5,$ $n=1,2,3$)
and of isoscalar $S$-wave processes $\pi\pi\to\pi\pi,K\overline{K},\eta\eta$~\cite{SBGKLN-prd18} also by
the description of $\gamma\gamma\to\pi^0\pi^0$ and of isoscalar $D$-wave four-channel $\pi\pi$ scattering
($\pi\pi\to\pi\pi,(2\pi)(2\pi),K\overline{K},\eta\eta$). Therefore, the parameters of isoscalar-scalar
and isoscalar-tensor resonances, obtained in our earlier analyses~\cite{SBKN-prd10}, remain unchanged.
We also intend to investigate the role of individual resonances (in the first turn the isoscalar-tensor
mesons) in the saturation of the energy spectrum of $\gamma\gamma\to\pi^0\pi^0$.

We first outline our model-independent $S$-matrix approach based on first principles,
such as analyticity and unitarity~\cite{SBKN-prd10}. It was used in our analysis of isoscalar $S$-
and $D$-wave $\pi\pi$ scattering. This approach had the most success for the $S$-wave scattering in
the two- and three-channel cases, because 4- and 8-sheeted Riemann surfaces, on which the corresponding
$S$-matrices are determined, respectively, are transformed onto the uniformization plane by a conformal
mapping (in three-channel case by neglecting the $\pi\pi$-threshold branch point~\cite{SBL-prd12}).
With $S=S_{bgr}S_{res}$, where $S_{bgr}$ is the background part, $S_{res}$ represents the contribution
of resonances which  is parameterized on the uniformization plane by poles and corresponding zeros
without additional assumptions~\cite{KMS-96}. For the parameterisation of $S_{res}$ by poles and zeros
it is convenient to use the Le Couteur-Newton relations~\cite{LeCou}:
\begin{eqnarray} \label{LCN}
S_{ii}&=&\frac{d(k_1,\cdots,k_{i-1},-k_i,k_{i+1},\cdots,k_N)}{d(k_1,\cdots,k_N)},\\
S_{ii}S_{jj}-S_{ij}^2&=&\frac{d(k_1,\cdots,k_{i-1},-k_i,k_{i+1},\cdots,k_{j-1},-k_j,k_{j+1},
\cdots,k_N)}{d(k_1,\cdots,k_N)}.\nonumber
\end{eqnarray}
The Jost matrix determinant $d(k_1,\cdots,k_N)$, being a real analytic function (i.e.
$d(s^*)=d^*(s)~{\rm for~all}~s$), has the only square-root branch-points at the channel momenta
$k_{i}=0$.

For $S$-wave $\pi\pi$ scattering, when using the uniformizing variable~\cite{SBL-prd12}:
\begin{equation}\label{univar}
w=\frac{\sqrt{(s-s_2)s_3} + \sqrt{(s-s_3)s_2}}{\sqrt{s(s_3-s_2)}}~~~~(s_2=4m_K^2 ~ {\rm and}~
s_3=4m_\eta^2)
\end{equation}
where we have neglected the $\pi\pi$-threshold branch point and allowed for the $K\overline{K}$-
and $\eta\eta$-threshold branch points and left-hand branch point at $s=0$ related
to the crossed channels, -- the Jost matrix determinant $d(k_1,k_2,k_3)$ is transformed
into the branch-point-free function $d(w)$. Parametrization of the $d(w)$ by poles and zeros,
representing resonances, and the analysis of three-channel $\pi\pi$ scattering (the result
has $\chi^2/\mbox{ndf}\approx1.16$) can be found in Refs.~\cite{SBLKN-jpgnpp14,SBL-prd12}.
Also note, that with the uniformizing variable~(\ref{univar}) it is impossible, in principle,
to describe a small near-$\pi\pi$-threshold region in the phase shift of the $\pi\pi$-scattering
amplitude. Therefore, to allow for 10 high-statistics data points for the $\pi\pi$ phase shift
in this region from the NA48/2 Collaboration~\cite{NA48/2}, we have continued the phase shift
to the $\pi\pi$-threshold as follows
\begin{equation}\label{totpipi}
\delta_{11}(s)=ArcSin\Bigl[2\sqrt{1-4m_{\pi^+}^2/s}\left(a_{\pi\pi} +
b_{\pi\pi}\frac{s-4m_{\pi^+}^2}{4m_{\pi^+}^2}\right)\Bigr]\theta(m_0^2-s) +
\overline{\delta_{11}}(s)\theta(s-m_0^2).
\end{equation}
We use the following set of the parameters. The parameters $a_{\pi\pi}$ and $b_{\pi\pi}$ are
fixed as $a_{\pi\pi}=0.282$ and $b_{\pi\pi}=0.222$.
$m_0 = 0.4115$ GeV is the scale parameter splitting the regions of
the $s$ variable into two parts: (1) small near-$\pi\pi$ threshold region
and (2) region where the phase shift of the $\pi\pi$-scattering amplitude
$\overline{\delta_{11}}(s)$ was obtained in our earlier model-independent
analysis~\cite{SBLKN-jpgnpp14,SBL-prd12} without allowing for the
near-$\pi\pi$-threshold data~\cite{NA48/2}; then $m_0$ is determined by
the smooth sewing of the indicated parts of the phase shift. With the phase
shift~(\ref{totpipi}) the satisfactory description of the isoscalar $S$-wave
$\pi\pi$-scattering (modulus and phase shift of the amplitude)
is with $\chi^2/\mbox{n.d.f.}\approx1.08$, and of the isoscalar $S$-wave
three-channel $\pi\pi$-scattering, i.e. of three coupled $S$-wave channels
$\pi\pi\to\pi\pi,K\overline{K},\eta\eta$, with $\chi^2/\mbox{n.d.f.}\approx1.12$.

When considering isoscalar $D$-wave multichannel $\pi\pi$-scattering, we are forced to deal at
least with four coupled channels $\pi\pi\to\pi\pi,(2\pi)(2\pi),K\overline{K},\eta\eta$, therefore,
our method of the uniformizing variable cannot be applied. Here we generated the resonance poles by
some four-channel Breit-Wigner forms in the Jost matrix determinant $d(k_1,k_2,k_3,k_4)=d_B d_{res}$,
where the resonance part is
\begin{equation}
d_{res}(s)=\prod_{r}
\biggl[M_r^2-s-i\sum_{j=1}^4\rho_{rj}^5R_{rj}f_{rj}^2\biggr] \,. 
\end{equation}
Here $\rho_{rj}=2k_j/\sqrt{M_r^2-4m_j^2}$, $f_{rj}^2/M_r$ 
is the partial width and $k_j=\sqrt{s-4m^2_j}/2$ is the j-channel momentum 
with the channel mass $m_j$, where 
$j= 1, 2, 3$, and $4$ denotes the $\pi\pi$, $(2\pi)(2\pi)$, $K\overline{K}$, 
and $\eta\eta$ channels, respectively. More datails can be found in Ref.~\cite{SBKN-prd10}.
For detailed description of the Breit-Wigner method we refer to paper~\cite{Bydzovsky:2016vdx}.

The Blatt--Weisskopf barrier factor for a tensor particle is
\begin{equation}
R_{rj}=\frac{9+\frac{3}{4}
\Bigl(\sqrt{M_r^2-4m_j^2}~r_{rj}\Bigr)^2+\frac{1}{16}\Bigl(\sqrt{M_r^2-4m_j^2}~r_{rj}\Bigr)^4}
{9+\frac{3}{4}\Bigl(\sqrt{s-4m_j^2}~r_{rj}\Bigr)^2+\frac{1}{16}\Bigl(\sqrt{s-4m_j^2}~r_{rj}\Bigr)^4},
\end{equation}
with radii $r_{rj}$ of 0.943 fm for all resonances in all channels except for $f_2(1270)$ and $f_2(1960)$.
In particular, for $f_2(1270)$: 1.498, 0.708, and 0.606 fm in the channels $\pi\pi$, $K\overline{K}$,
and $\eta\eta$, respectively; for $f_2(1960)$: 0.296 fm in the channel $K\overline{K}$).
The description of the accessible data on isoscalar $D$-wave four-channel $\pi\pi$ scattering
($\pi\pi\to\pi\pi,(2\pi)(2\pi),K\overline{K},\eta\eta$)~\cite{Hya73,Lin92} results
in $\chi^2/\mbox{n.d.f.}\approx 1.58$.

\section{Numerical results}

With the obtained amplitudes of isoscalar $S$- and $D$-wave multi-channel $\pi\pi$-scattering
of Eqs.~(\ref{S-D-ampl}) and~(\ref{S-D-ampl2}) we also obtain a satisfactory description of the
cross sections for $\gamma\gamma\to\pi^0\pi^0$. The fit to the data from the Crystal Ball~\cite{CrBall_gg}
and Belle~\cite{Belle13,Belle_gg} Collaborations in the energy region from the $\pi\pi$ threshold to
$\approx2.23$~GeV has a $\chi^2$, averaged on the number of experimental points, of
$\chi^2/\mbox{n.p.}\approx 1.30$.

The free parameters in Eqs.~(\ref{contrib}) are found to be
$\alpha_{10}=-4.39948$, $\gamma_1=-2.414$, $\alpha_{20}=56.97042$, $\alpha_{30}=-13.64394$;
$\beta_{20}=0.33308$, $\gamma_2=0.02913$, $\beta_{21}=-150.47327$, $\beta_{22}=148.66336$;
$\beta_{31}=-121.01348$, $\beta_{32}=93.55851$;
$\delta_{10}=320.91310$, $\rho_1=0.075394$, $\delta_{11}=720.63174$, $\delta_{12}=-166.62981$;
$\delta_{20}=35.65701$, $\rho_2=0.075394$, $\delta_{21}=182.68994$, $\delta_{22}=-42.44117$;
$\delta_{31}=31.15567$, $\delta_{32}=-6.46263$;
$\delta_{41}=783.32892$, $\delta_{42}=-200.53452$.
Note that, e.g., the parameters $\delta_{10}=320.91310$ and $\delta_{20}=35.65701$,
related to isovector and isoscalar components of the hadron electromagnetic current~(\ref{MVD}),
respectively, satisfy approximately the corresponding relation~(\ref{MVDrelation}).

Considering (with the determined parameters) the quantities $a(\gamma\gamma\to n)$ and
$b(\gamma\gamma\to n)$, which describe transitions to the final $\pi^0\pi^0$ state through
the corresponding intermediate states $n=\pi^+\pi^-,2\pi^+2\pi^-,K^+K^-,\eta\eta$,
one can conclude, that the contribution of isoscalar $D$-wave multi-channel
$\pi\pi$ scattering is dominant in comparison to the $S$-wave one.
This could be expected when considering the data for the energy spectrum with
a large enhancement in the $f_2(1270)$ region.
\begin{figure}[!hb]
\begin{center}
\includegraphics[width=0.495\textwidth,angle=0]{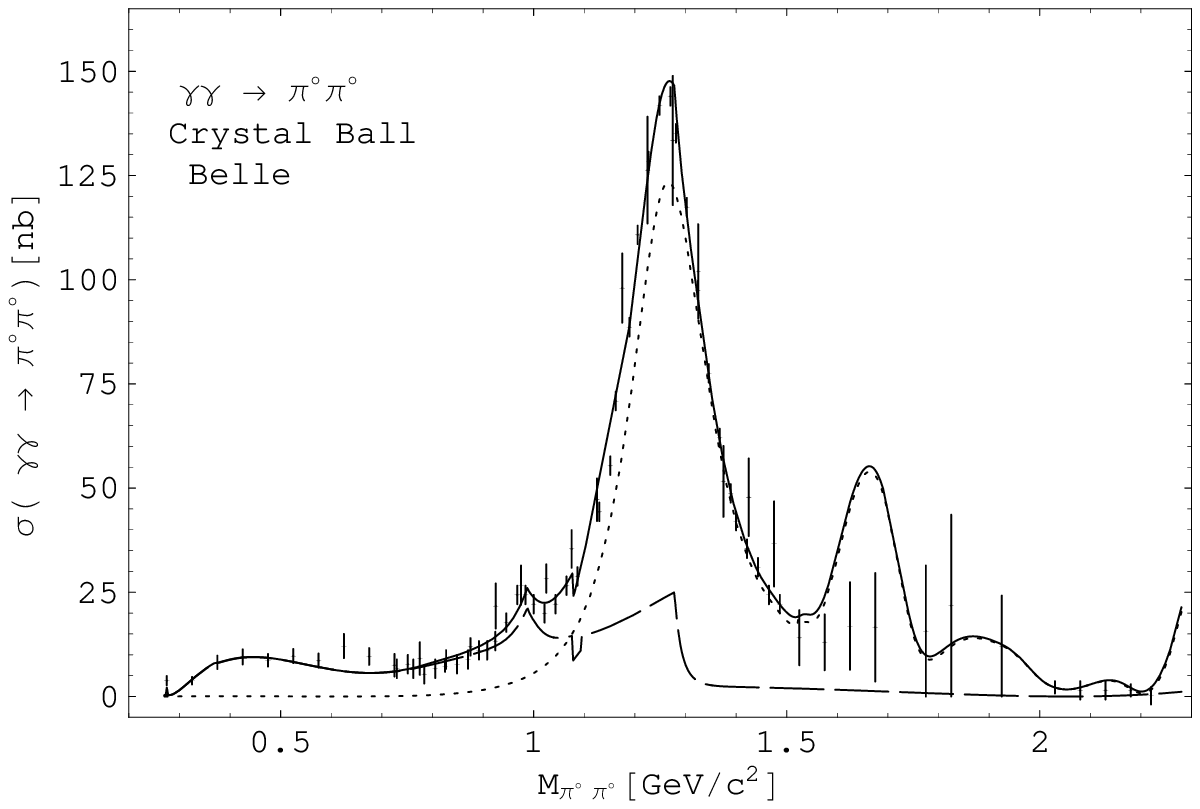}
\includegraphics[width=0.495\textwidth,angle=0]{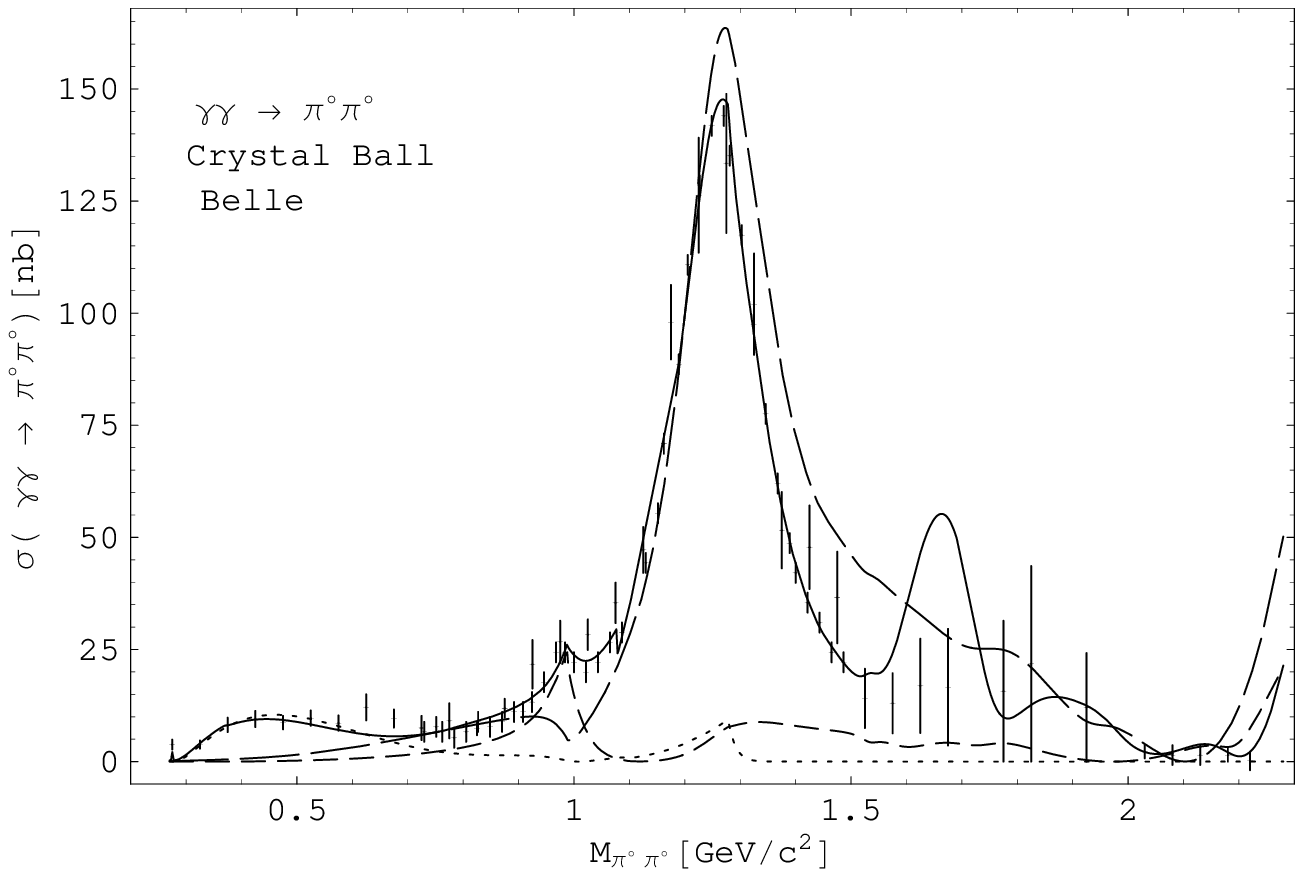}
\hspace*{1cm}(a) \hspace*{7.5cm} (b)
\caption{Description of the cross section of the $\gamma\gamma\to\pi^0\pi^0$ process.
The experimental data are taken from the Crystal Ball~\cite{CrBall_gg} and
Belle~\cite{Belle13,Belle_gg} Collaborations. \\
(a): Full result for the cross section (solid line) is compared with
contributions of the $S$-wave (long-dashed line) and $D$-wave (dotted line). \\
(b): Full result for the cross section (solid line) is compared with
contributions of the coupled channels to the calculated energy spectrum
by the $\pi\pi$ channel (long-dashed line); $K\overline{K}$ channel (dotted line);
$\eta\eta$ channel (short-dashed line).}
\label{Partialcontr}
\end{center}
\end{figure}
In Fig.~\ref{Partialcontr} we present our description of data on the energy spectrum
of cross section of $\gamma\gamma\to\pi^0\pi^0$ from the Crystal Ball~\cite{CrBall_gg}
and Belle~\cite{Belle13,Belle_gg} Collaborations. In Fig.~\ref{Partialcontr}(a) we compare
full results (solid line) with separate contributions from the $S$ (long-dashed line)-
and $D$ (dotted line)-waves. Note, the energy spectrum of $\gamma\gamma\to\pi^0\pi^0$ from
the $\pi\pi$-threshold to the $K\overline{K}$-threshold is almost completely determined
by the $S$-wave and above the $K\overline{K}$-threshold (and especially above
$\approx1.3$~GeV) by the $D$-wave contribution. The observed bell-shaped behaviour of
the energy spectrum in the near-$\pi\pi$-threshold region is related to the $S$-wave
contribution, whereas dips and structures above 1.3~GeV are due to the $f_2$-resonances
contributing to the $D$-wave including their interference.
In Fig.~\ref{Partialcontr}(b) we display the comparison of full result (solid line) with
contributions of the individual coupled channels --- $\pi\pi$ channel (long-dashed line),
$K\overline{K}$ channel (dotted line), $\eta\eta$ channel (short-dashed line).
One should stress that the bell-shaped behaviour of the near-$\pi\pi$-threshold energy spectrum
is mainly related to the $S$-wave $K\overline{K}$ intermediate state contribution.
We also observe a sizable contribution of the $D$-wave process
$\gamma\gamma\to\pi^+\pi^-\to\pi^0\pi^0$. Above $\approx0.85$~GeV we also get
a noticeable contribution of the $D$-wave process $\gamma\gamma\to\eta\eta\to\pi^0\pi^0$
(in comparison with the ones of the other coupled processes
$\gamma\gamma\to K^+K^-\to\pi^0\pi^0$ and
$\gamma\gamma\to(\pi^+\pi^-)(\pi^+\pi^-)\to\pi^0\pi^0$). The large contribution of
the one-pion exchange to the crossing channel of the $D$-wave process
$\gamma\gamma\to\pi\pi\to\pi\pi$ is expected. The intermediate state $\eta\eta$ in
the reaction $\gamma\gamma\to\pi^0\pi^0$ implies, of course, a preceding pair of
charged particles ($\pi^+\pi^-$) in the intermediate state, that is, the process
$\gamma\gamma\to\pi^+\pi^-\to\eta\eta\to\pi^0\pi^0$ and the one-pion exchange between
the $\gamma\pi^+$ and $\gamma\pi^-$ vertices. In any case, this explains the noticeable
contribution of the $D$-wave process $\gamma\gamma\to\eta\eta\to\pi^0\pi^0$.

From Fig.~\ref{Partialcontr} it is evident that data of the Crystal Ball~\cite{CrBall_gg}
and Belle~\cite{Belle13,Belle_gg} Collaborations are quite consistent.
Though the Belle Collaboration data are only available from 0.73 to 1.49~GeV,
the Crystal Ball data reside in the energy range from the $\pi\pi$~threshold up to
$\approx2.22$~GeV, however they have rather big statistical errors. It is also seen that
there is some discrepancy between our calculations and the experimental energy spectrum
in interval from $\approx1.5$ to $\approx1.73$~GeV. Therefore, keeping in mind the uncertain
status of a number of isoscalar-tensor mesons (six among twelve mesons, indicated in
the PDG issue~\cite{PDG}, need confirmation), it is worth considering the role of individual
resonances (being situated in the interval from 1.5 to 1.73~GeV) in the saturation of
the energy spectrum of $\gamma\gamma\to\pi^0\pi^0$.
These are the resonances $f_2(1534.7)$, $f_2(1601.5)$, $f_2(1719.8)$, and $f_2(1760)$.

We would like to stress, that the obtained quite satisfactory description of data on the energy
spectrum of $\gamma\gamma\to\pi^0\pi^0$ cross section jointly with the satisfactory combined
description of data on the isoscalar $S$-wave three-channel $\pi\pi$ scattering
($\pi\pi\to\pi\pi,K\overline{K},\eta\eta$), on the isoscalar $D$-wave four-channel $\pi\pi$
scattering ($\pi\pi\to\pi\pi,(2\pi)(2\pi),K\overline{K},\eta\eta$) and on the above-indicated
decays of charmonia and bottomonia~\cite{SBGKLN-prd18} successfully confirms all our preceding
results on scalar mesons obtained in Ref.~\cite{SBLKN-prd14}).

Next we consider the role of individual isoscalar-tensor mesonic resonances in saturation of
the energy spectrum of the process $\gamma\gamma\to\pi^0\pi^0$. It is worth making, because
interval $1.5-1.73$~GeV, where our calculations diverge with the data
[see Fig.~\ref{Partialcontr}(b)], are in the region described mainly by the $D$-wave.

Further, switching off the $f_2$ resonances, grouped around the energy interval $1.5-1.73$~GeV,
we have performed a combined analysis (for each case) of $\gamma\gamma\to\pi^0\pi^0$,
of the isoscalar $S$-wave processes $\pi\pi\to\pi\pi,K\overline{K},\eta\eta$, and of the isoscalar
$D$-wave processes $\pi\pi\to\pi\pi,(2\pi)(2\pi),K\overline{K},\eta\eta$.
In Table~\ref{PDG} we make correspondence of the omitted resonances 
in the Particle Data Group (PDG)~\cite{PDG} with our predictions. In the following we use 
the notations for the resonances $f_J(M_{\rm our})$, where $M_{\rm our}$ is our prediction 
for the mass of the corresponding state. 
In Table~\ref{tab:khi_sqrd} we list $\chi^2/\mbox{n.p.}$, for the former description 
of decay processes and the total $\chi^2/\mbox{n.d.f.}$, calculated for the combined 
analyses also for each case.

\begin{table}[htb!]
\caption{Correspondence of the omitted $f_2$ resonances in the PDG~\cite{PDG} 
with our predictions} 
\begin{center}
\begin{tabular}{|c|c|c|c|c|c|c|} 
\hline
PDG~\cite{PDG}   & $f_2(1430)$   & $f_2(1565)$   & $f_2(1640)$   & $f_2(1810)$   & $f_2(1910)$   & $f_2(2150)$  \\ 
\hline 
Our predictions  & $f_2(1450.5)$ & $f_2(1534.7)$ & $f_2(1601.5)$ & $f_2(1719.8)$ & $f_2(1760.0)$ & $f_2(2202.0)$ \\ 
\hline 
\end{tabular}
\label{PDG}
\end{center}

\vspace*{.5cm}

\caption{$\chi^2$ analysis}
\begin{center}
{
\hspace*{-1.5cm}
\begin{tabular}{|c|c|c|c|c|c|c|c|} \hline
{Omitted state} & -- & $f_2(1450.5)$ & $f_2(1534.7)$ & $f_2(1601.5)$ & $f_2(1719.8)$ & $f_2(1760.0)$ & $f_2(2202.0)$  
\\ \hline 
$\chi^2/\mbox{n.p.}$         & 1.30 & 1.50 & 1.51 & 1.47 & 1.53 & 1.51 & 1.52 \\ \hline
total $\chi^2/\mbox{n.d.f.}$ & 1.32 & 1.51 & 1.36 & 1.51 & 1.58 & 1.48 & 1.44 \\ \hline\hline
{Omitted}            & $f_2(1430)$~\& & $f_2(1534.7)$~\& & $f_2(1534.7)$~\& & $f_2(1534.7)$~\& & $f_2(1601.5)$~\& & $f_2(1601.5)$~\& & $f_2(1719.8)$~\& \\
{states}             & $f_2(1534.7)$   & $f_2(1601.5)$    & $f_2(1719.8)$    & $f_2(1760)$    & $f_2(1719.8)$    & $f_2(1760)$    & $f_2(1760)$ \\ \hline
$\chi^2/\mbox{n.p.}$     & 1.56            & 1.35             & 1.51             & 1.53             & 1.17             & 1.47            & 1.22 \\ \hline
total $\chi^2/\mbox{n.d.f.}$ & 1.52        & 1.54             & 1.58             & 1.48             & 1.81             & 1.61            & 1.60 \\ \hline
\end{tabular}}
\label{tab:khi_sqrd}
\end{center}
\end{table}
\begin{figure}[!thb]
\begin{center}
\includegraphics[width=0.495\textwidth,angle=0]{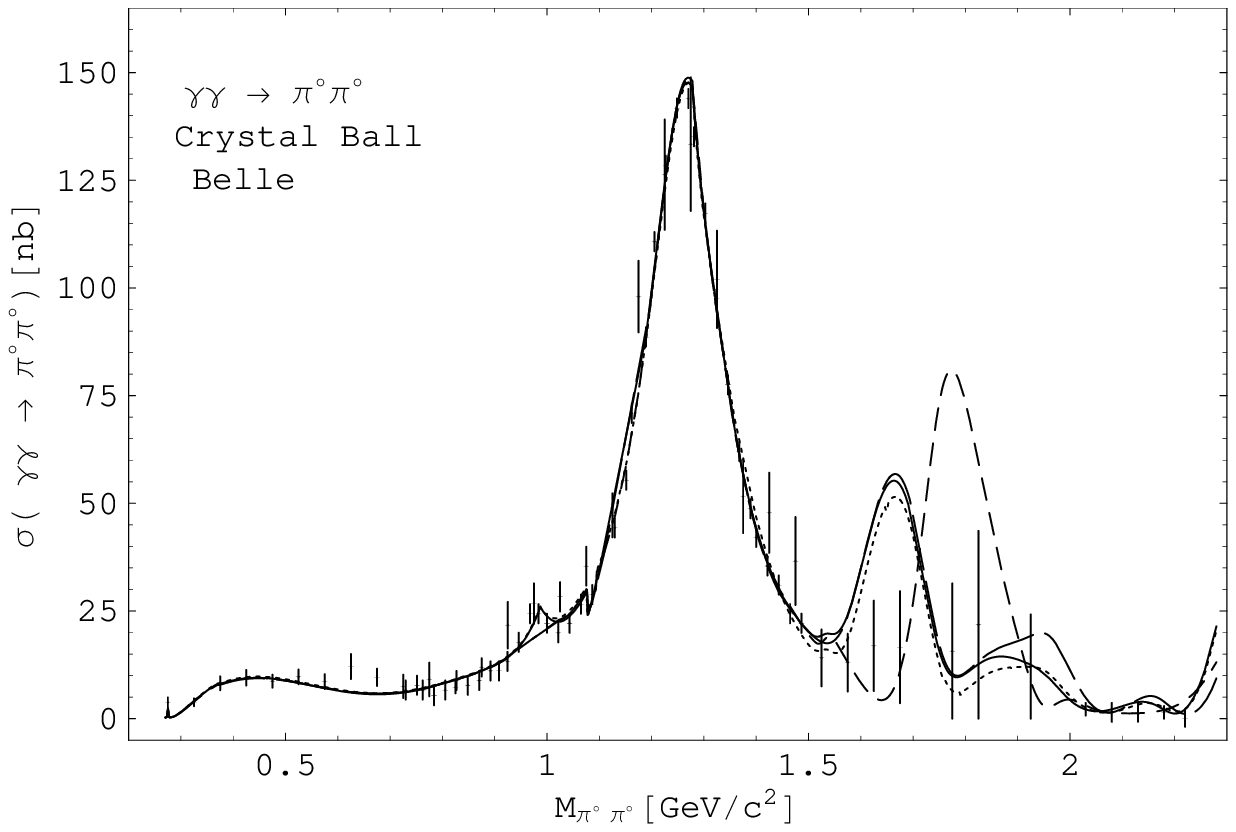}
\includegraphics[width=0.495\textwidth,angle=0]{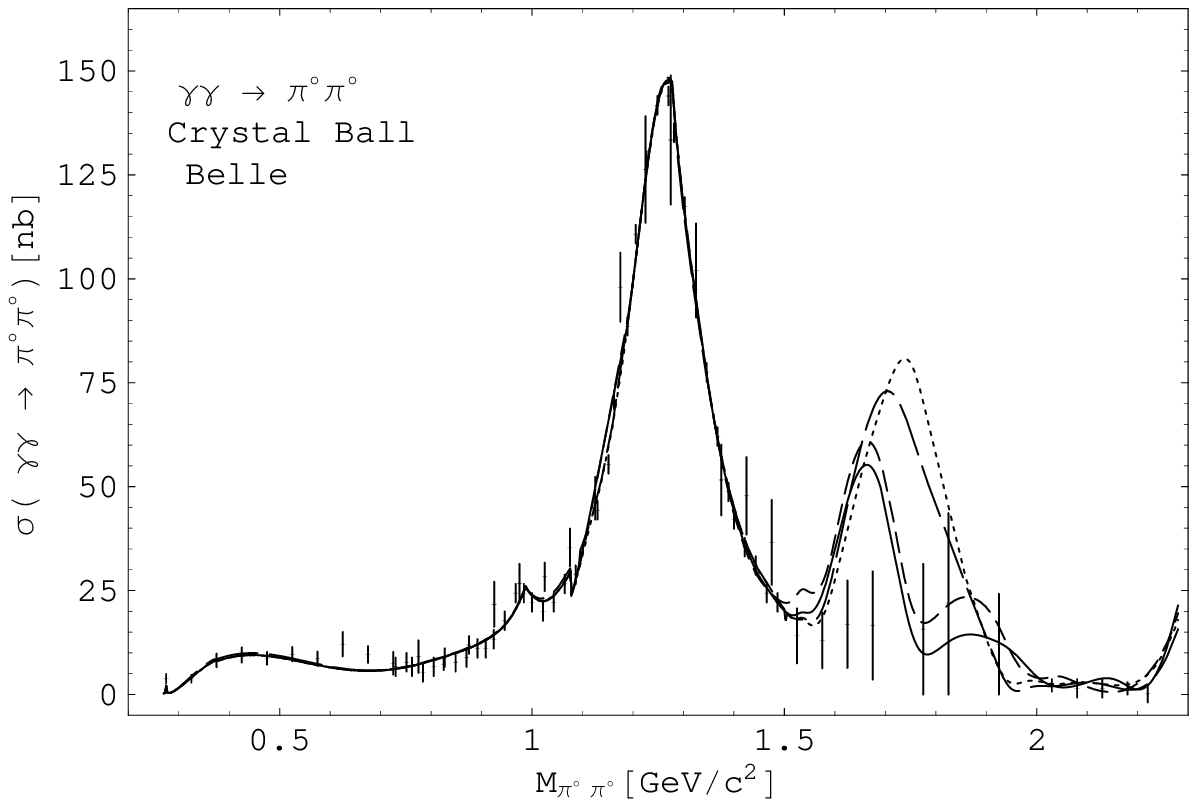}\\
\hspace*{1cm}(a) \hspace*{7.5cm} (b) \\[2mm]
\includegraphics[width=0.495\textwidth,angle=0]{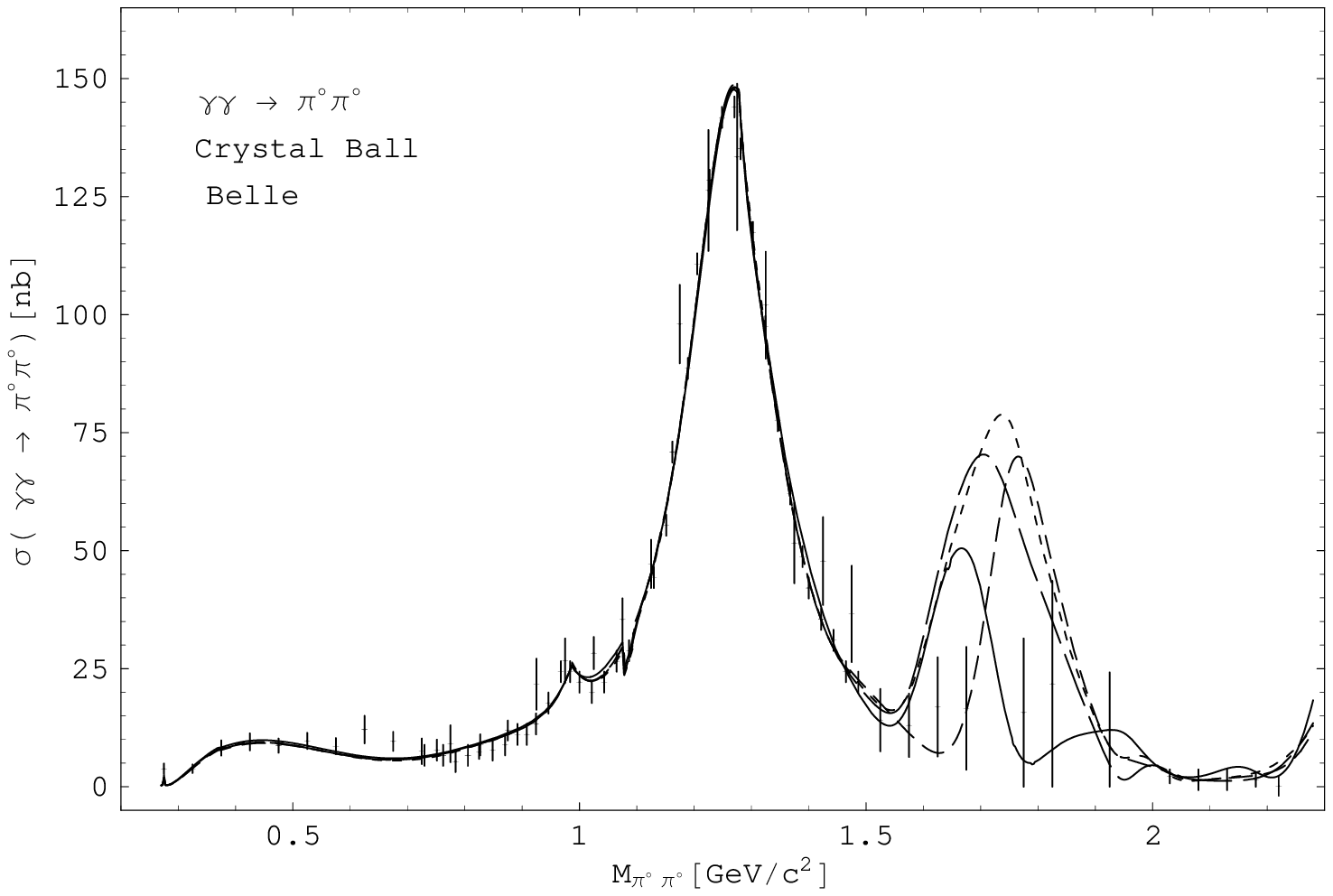}
\includegraphics[width=0.495\textwidth,angle=0]{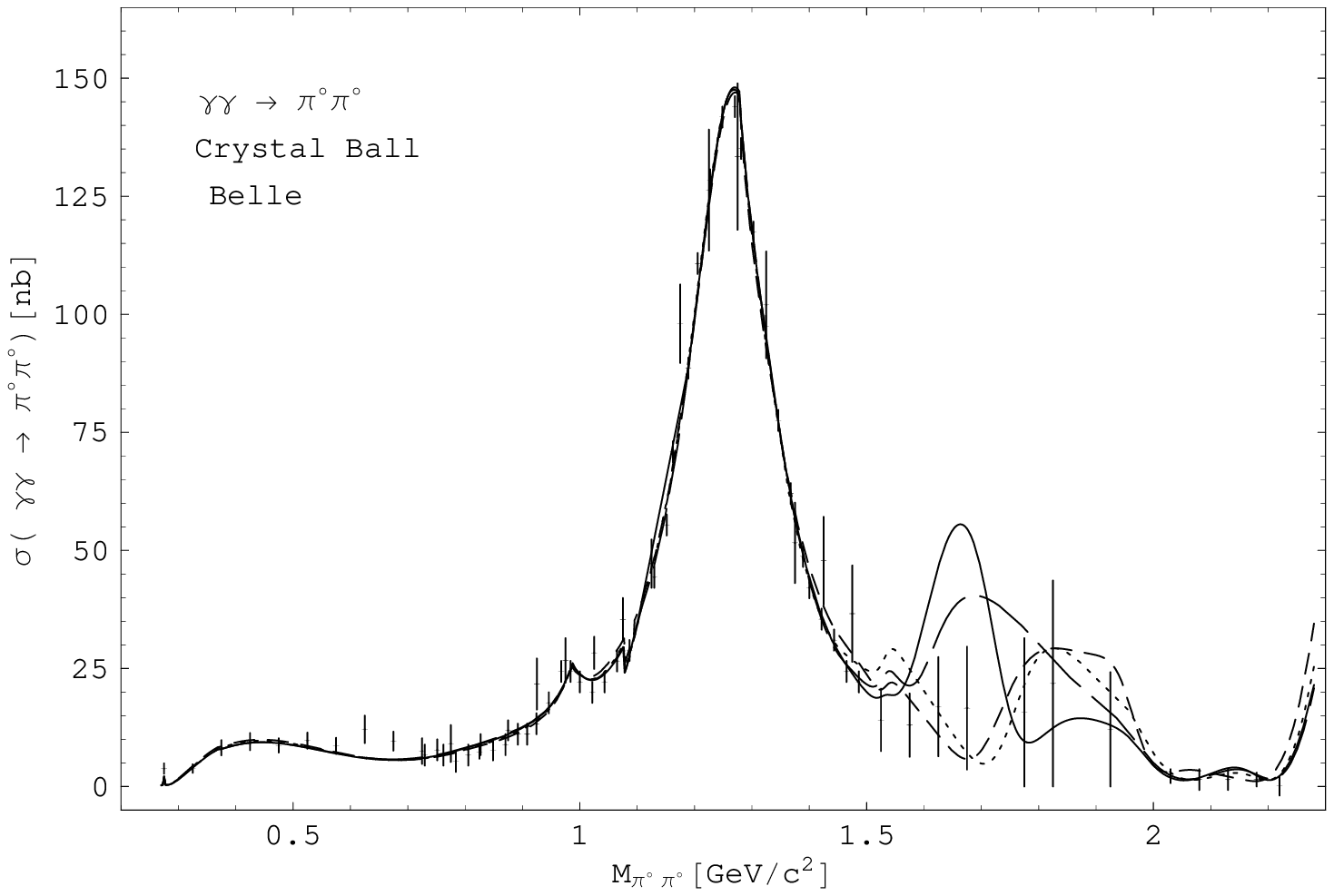}
\hspace*{1cm}(c) \hspace*{7.5cm} (d) \\
\caption{Results for the cross section of the $\gamma\gamma\to\pi^0\pi^0$ for different scenarios
when relevant tensor mesons ($f_2$-resonances) are switched off from the analysis.\\
(a): Comparison of the scenarios:
(1) full result with taking into account
of all involved $f_0$- and $f_2$-resonances (solid line);
(2) $f_2(1450.5)$ state is excluded (dotted line);
(3) $f_2(1601.5)$ state is excluded (short-dashed line);
(4) $f_2(1534.7)$ state is excluded (long-dashed line). \\
(b): Comparison of the scenarios:
(1) full result with taking into account
of all involved $f_0$- and $f_2$-resonances (solid line);
(2) $f_2(1719.8)$ state is excluded (dotted line);
(3) $f_2(2202)$ state is excluded (short-dashed line);
(4) $f_2(1760)$ state is excluded (long-dashed line). \\
(c): Comparison of the scenarios:
(1) $f_2(1450.5)$ and $f_2(1534.7)$ states are excluded (solid line);
(2) $f_2(1534.7)$ and $f_2(1719.8)$ states are excluded (dotted line);
(3) $f_2(1534.7)$ and $f_2(1601.5)$ states are excluded (short-dashed line);
(4) $f_2(1534.7)$ and $f_2(1760)$ states are excluded (long-dashed line). \\
(d): Comparison of the scenarios:
(1) full result with taking into account
of all involved $f_0$- and $f_2$-resonances (solid line);
(2) $f_2(1601.5)$ and $f_2(1719.8)$ states are excluded (dotted line);
(3) $f_2(1601.5)$ and $f_2(1760)$ states are excluded (short-dashed line);
(4) $f_2(1719.8)$ and $f_2(1760)$ states are excluded (long-dashed line).}
\label{fig:fit}
\end{center}
\end{figure}
In Fig.~\ref{fig:fit} we show our predictions for the cross section of the
$\gamma\gamma\to\pi^0\pi^0$ process for different scenarios, when relevant
tensor mesons ($f_2$-resonances) are switched off from the analysis.
In particular, in Fig.~\ref{fig:fit}(a) we present the comparison of
the following scenarios:
(1) full result with taking into account
of all involved $f_0$- and $f_2$-resonances (solid line);
(2) $f_2(1450.5)$ state is excluded (dotted line);
(3) $f_2(1601.5)$ state is excluded (short-dashed line);
(4) $f_2(1534.7)$ state is excluded (long-dashed line).
In Fig.~\ref{fig:fit}(b) we make comparison of the scenarios:
(1) full result with taking into account
of all involved $f_0$- and $f_2$-resonances (solid line);
(2) $f_2(1719.8)$ state is excluded (dotted line);
(3) $f_2(2202)$ state is excluded (short-dashed line);
(4) $f_2(1760)$ state is excluded (long-dashed line).
Then, in Fig.~\ref{fig:fit}(c) we display the results for the scenarios:
(1) $f_2(1450.5)$ and $f_2(1534.7)$ states are excluded (solid line);
(2) $f_2(1534.7)$ and $f_2(1719.8)$ states are excluded (dotted line);
(3) $f_2(1534.7)$ and $f_2(1601.5)$ states are excluded (short-dashed line);
(4) $f_2(1534.7)$ and $f_2(1760)$ states are excluded (long-dashed line).
Finally, in Fig.~\ref{fig:fit}(d) we compare our predictions
for the following scenarios:
(1) full result with taking into account
of all involved $f_0$- and $f_2$-resonances (solid line);
(2) $f_2(1601.5)$ and $f_2(1719.8)$ states are excluded (dotted line);
(3) $f_2(1601.5)$ and $f_2(1760)$ states are excluded (short-dashed line);
(4) $f_2(1719.8)$ and $f_2(1760)$ states are excluded (long-dashed line).

The description of data in the energy region from $\pi\pi$~threshold to about 1.45~GeV (and in
a short region above 2.04~GeV) is almost the same when turning off the different $f_2$ contributions.
The main differences of the calculated energy dependencies of these variants are situated in the
interval from 1.45 to 2.04~GeV.

In Fig.~\ref{fig:fit} we see that in the interval 1.5--1.73 GeV, where the experimental energy
spectrum of the process $\gamma\gamma\to\pi^0\pi^0$ differs considerably from our calculations,
the data have rather big experimental errors. 
Therefore, the coupled-channel method described here is promising for
exclusion of particular f2 mesons once more data are available. 

The results in Table~\ref{tab:khi_sqrd} show that if the pairs of resonances 
$f_2(1601.5) \& f_2(1719.8)$ and $f_2(1719.8) \& f_2(1760)$ are omitted  
a better fit to the $\gamma\gamma\to\pi^0\pi^0$ data is obtained 
($\chi^2$/n.p. is smaller: 1.17 and 1.22, respectively) which indicates that these 
resonances are not important in describing the energy spectra of 
$\gamma\gamma\to\pi^0\pi^0$.
However, omitting these states results in an unsatisfactory 
description of the coupled-channel scattering  
$\pi\pi\to K\overline{K},\eta\eta$ ($\chi^2$/ndf is much larger) 
and therefore the indicated states cannot be excluded in the 
combined analysis. 

\begin{figure}[!thb]
\begin{center}
\includegraphics[width=0.495\textwidth,angle=0]{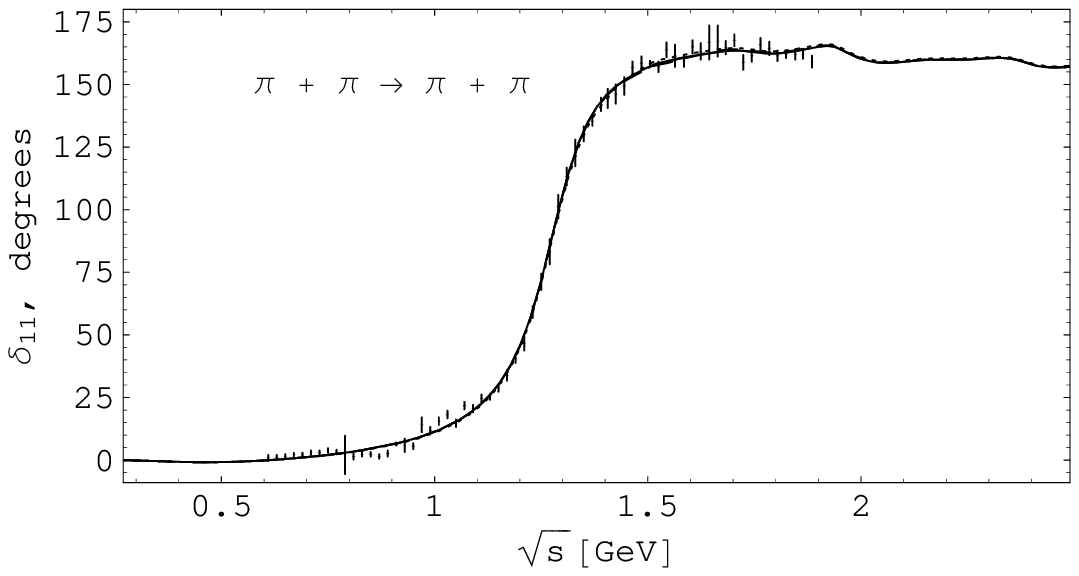}
\includegraphics[width=0.495\textwidth,angle=0]{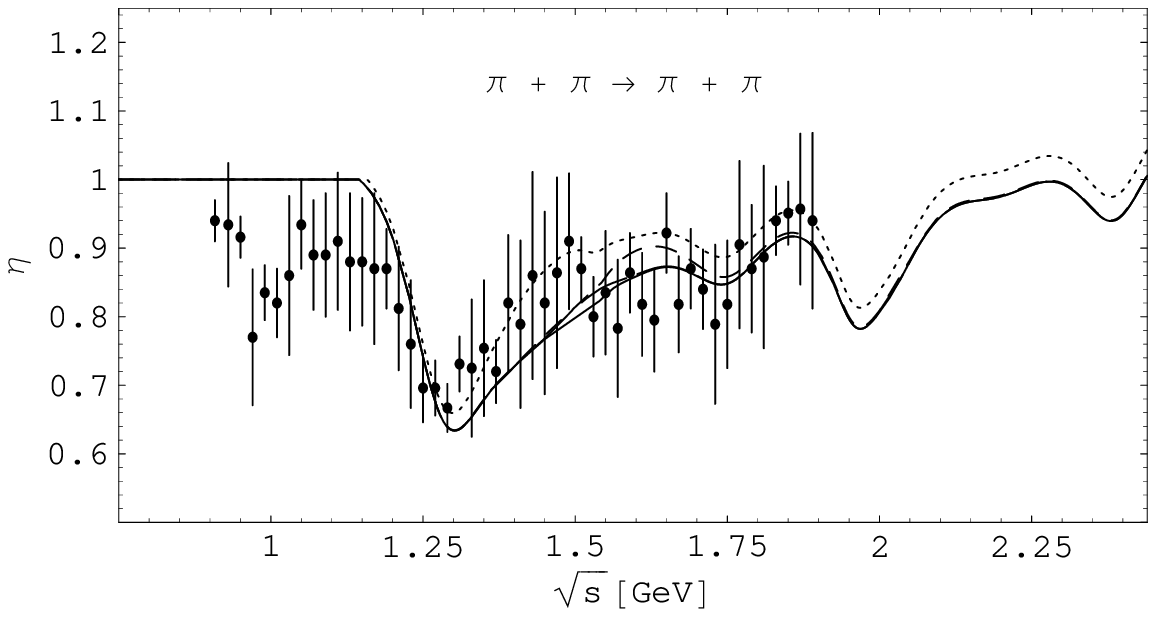}\\
\hspace*{1cm}(a) \hspace*{7.5cm} (b) \\[5mm]
\includegraphics[width=0.495\textwidth,angle=0]{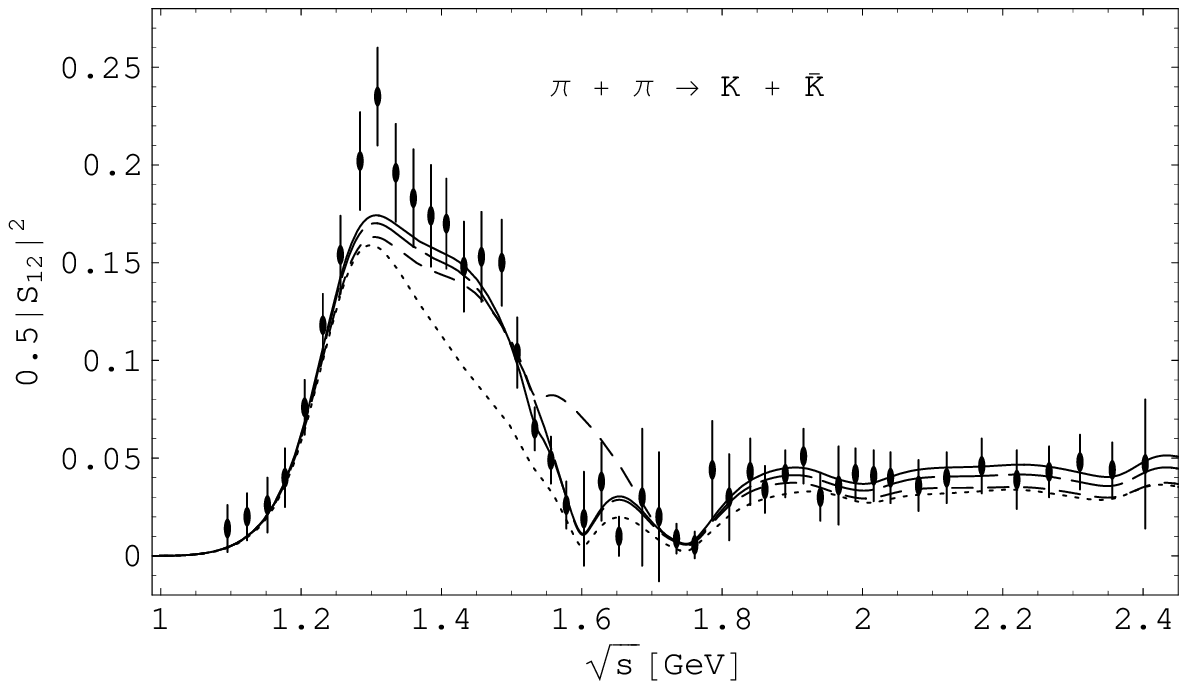}
\includegraphics[width=0.495\textwidth,angle=0]{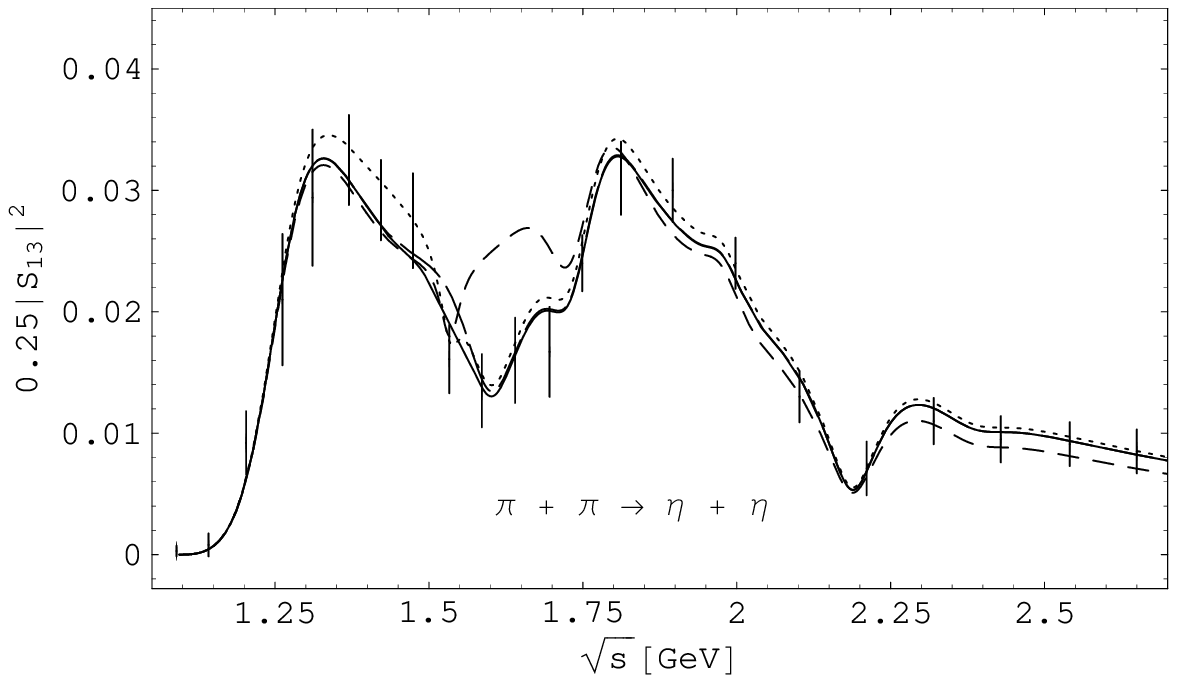}\\
\hspace*{1cm}(c) \hspace*{7.5cm} (d) \\
\caption{Scattering properties of the $\pi\pi \to \pi\pi, K\bar K, \eta\eta$ processes
for four scenarios:
(1) full result with inclusion of all relevant $f_2$-resonances (solid lines);
(2) $f_2(1450.5)$ state is excluded (dotted line);
(3) $f_2(1601.5)$ state is excluded (short-dashed line);
(4) $f_2(1534.7)$ state is excluded (long-dashed line). \\
(a): Phase shift and in the $D$-wave $\pi\pi$-scattering. \\
The experimental data are from Refs.~\cite{Hya73} and~\cite{Lin92}.\\
(b): Modulus of the $S$-matrix element in the $D$-wave $\pi\pi$-scattering. \\
(c): Squared moduli of the $S$-matrix element in the $D$-wave $\pi\pi \to K \bar K$ process.\\
(d): Squared moduli of the $S$-matrix element in the $D$-wave $\pi\pi \to \eta\eta$ process.}
\label{all_2_4_7}
\end{center}
\end{figure}

\begin{figure}[!thb]
\begin{center}
\includegraphics[width=0.495\textwidth,angle=0]{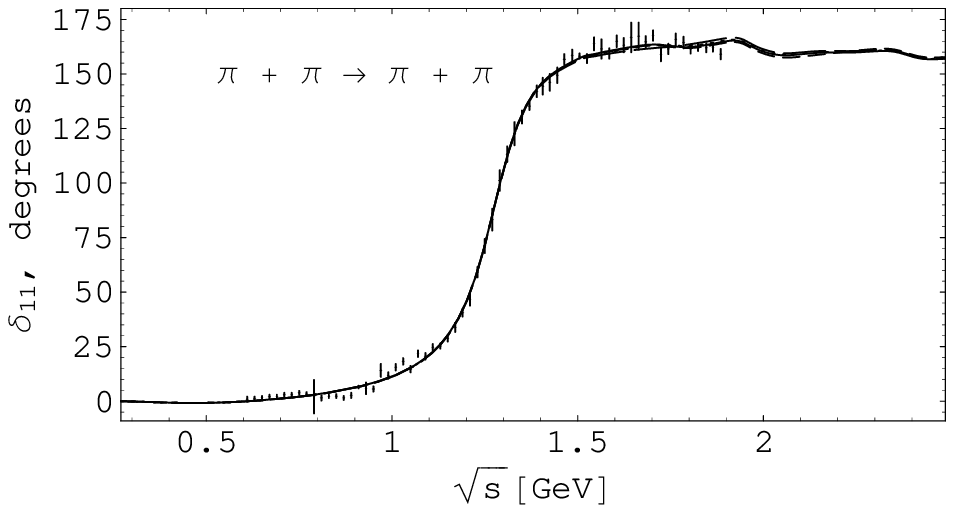}
\includegraphics[width=0.495\textwidth,angle=0]{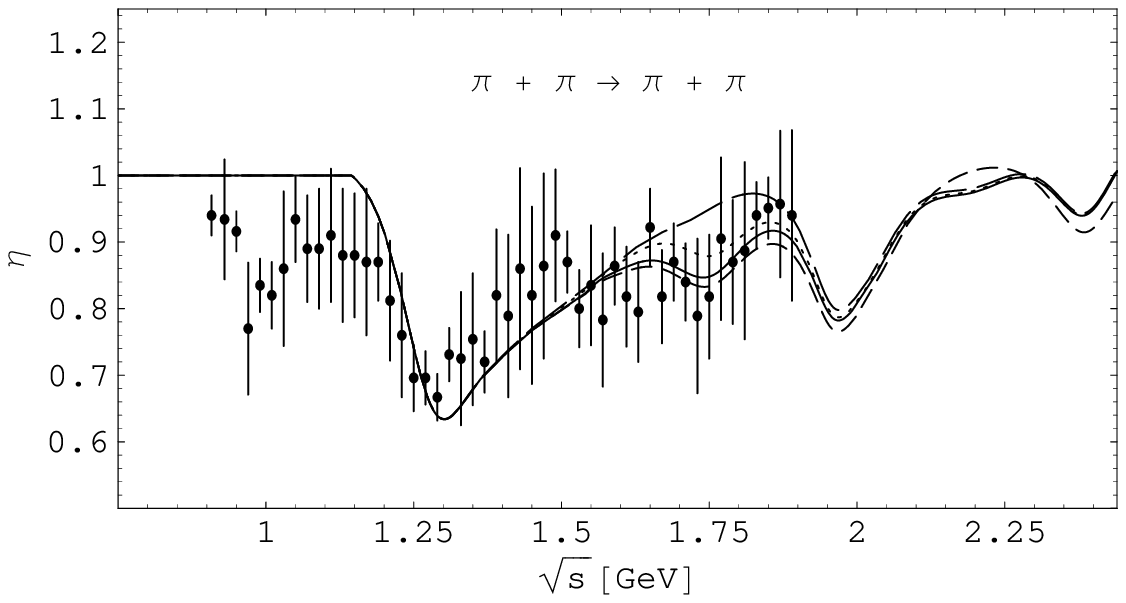}\\
\hspace*{1cm}(a) \hspace*{7.5cm} (b) \\[5mm]
\includegraphics[width=0.495\textwidth,angle=0]{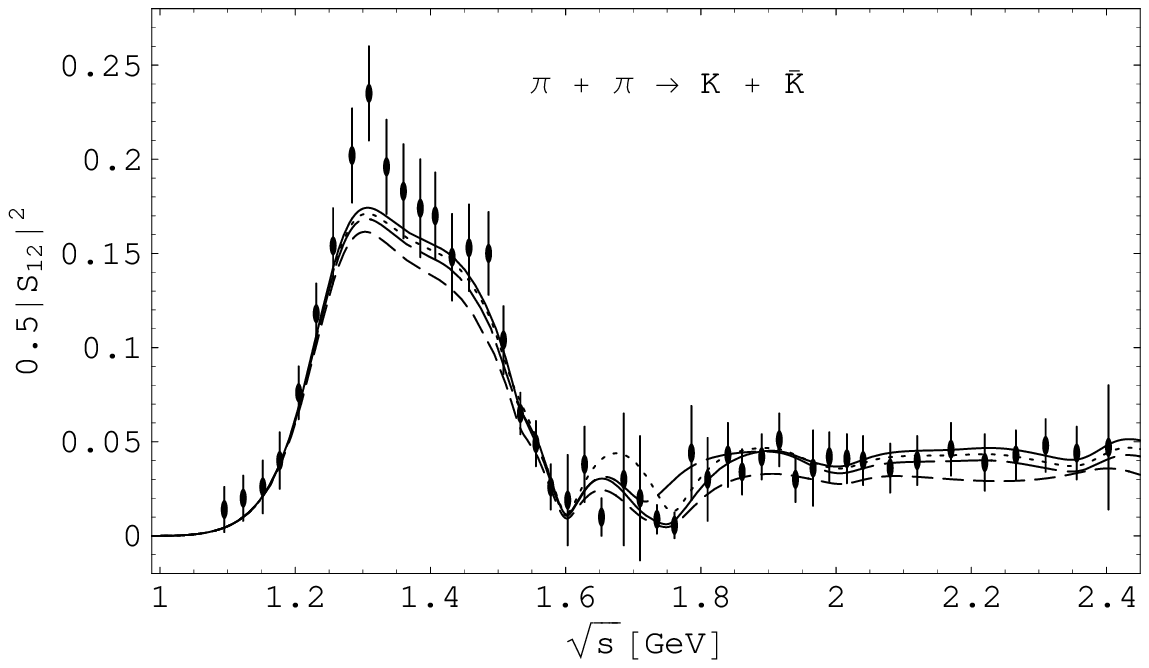}
\includegraphics[width=0.495\textwidth,angle=0]{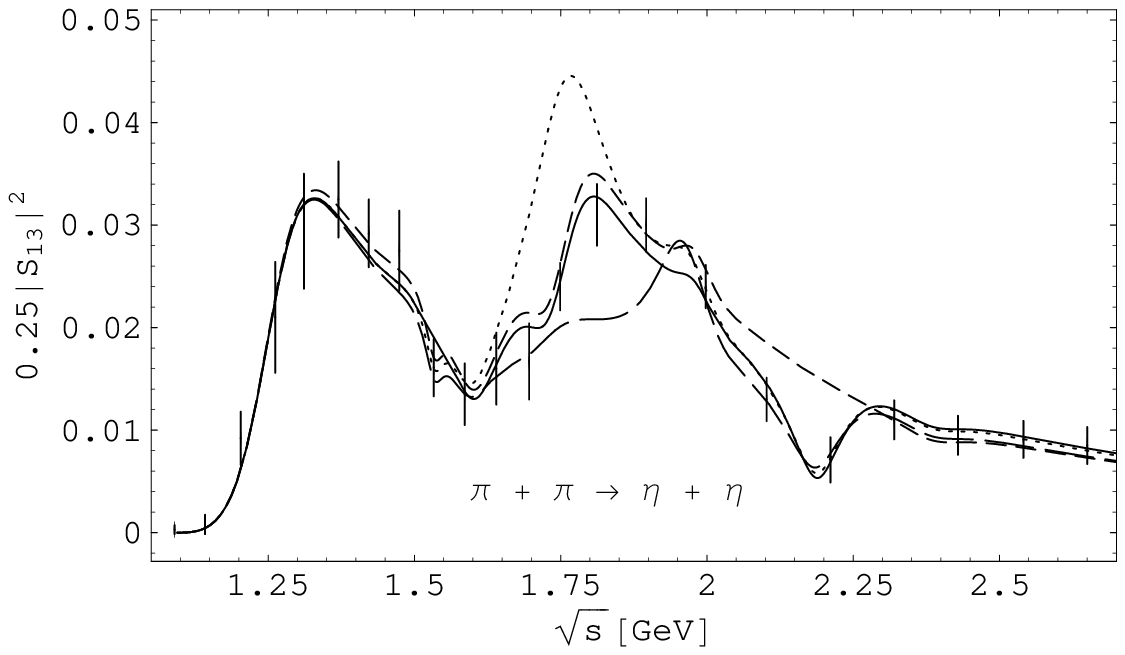}\\
\hspace*{1cm}(c) \hspace*{7.5cm} (d) \\
\caption{Scattering properties of the $\pi\pi \to \pi\pi, K\bar K, \eta\eta$ processes
for four scenarios:
(1) full result with inclusion of all relevant $f_2$-resonances (solid lines);
(2) $f_2(1719.8)$ state is excluded (dotted line);
(3) $f_2(2202)$ state is excluded (short-dashed line);
(4) $f_2(1760)$ state is excluded (long-dashed line). \\
The experimental data are from Refs.~\cite{Hya73} and~\cite{Lin92}.\\
(a): Phase shift and in the $D$-wave $\pi\pi$-scattering. \\
(b): Modulus of the $S$-matrix element in the $D$-wave $\pi\pi$-scattering. \\
(c): Squared moduli of the $S$-matrix element in the $D$-wave $\pi\pi \to K \bar K$ process.\\
(d): Squared moduli of the $S$-matrix element in the $D$-wave $\pi\pi \to \eta\eta$ process.}
\label{all_3_8_5}
\end{center}
\end{figure}

\begin{figure}[!thb]
\begin{center}
\includegraphics[width=0.495\textwidth,angle=0]{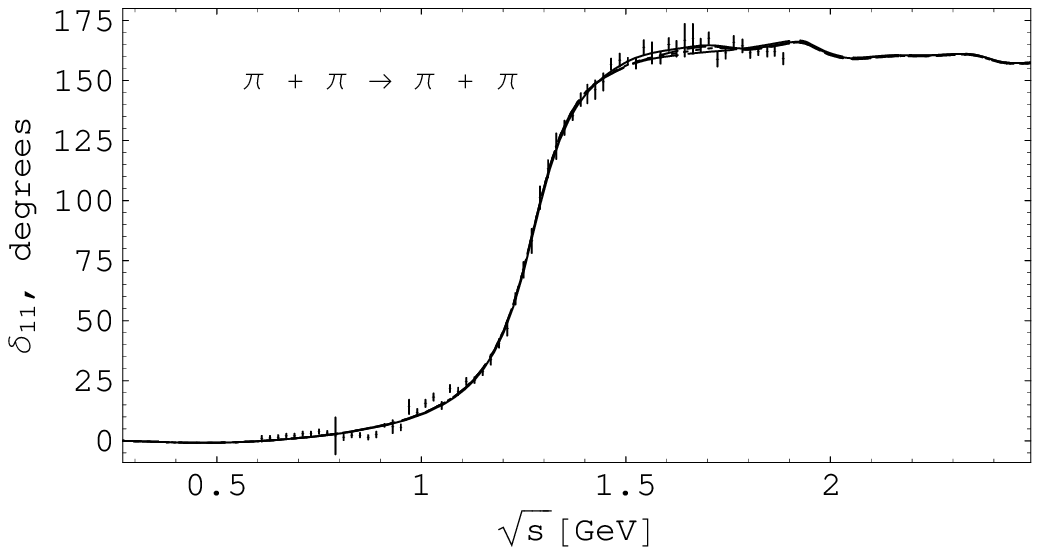}
\includegraphics[width=0.495\textwidth,angle=0]{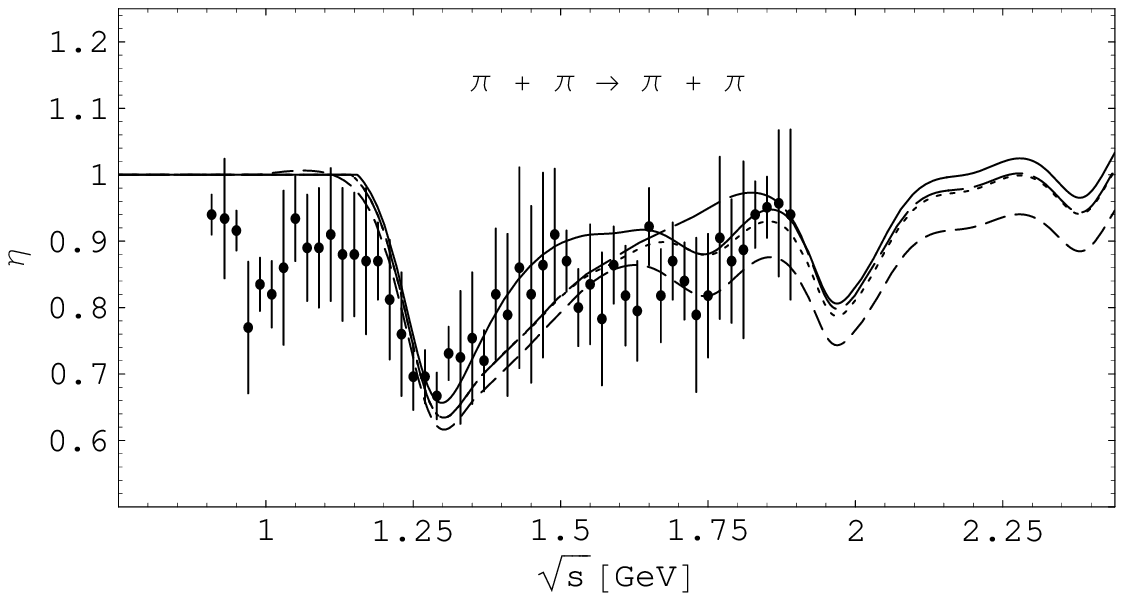}\\
\hspace*{1cm}(a) \hspace*{7.5cm} (b) \\[5mm]
\includegraphics[width=0.495\textwidth,angle=0]{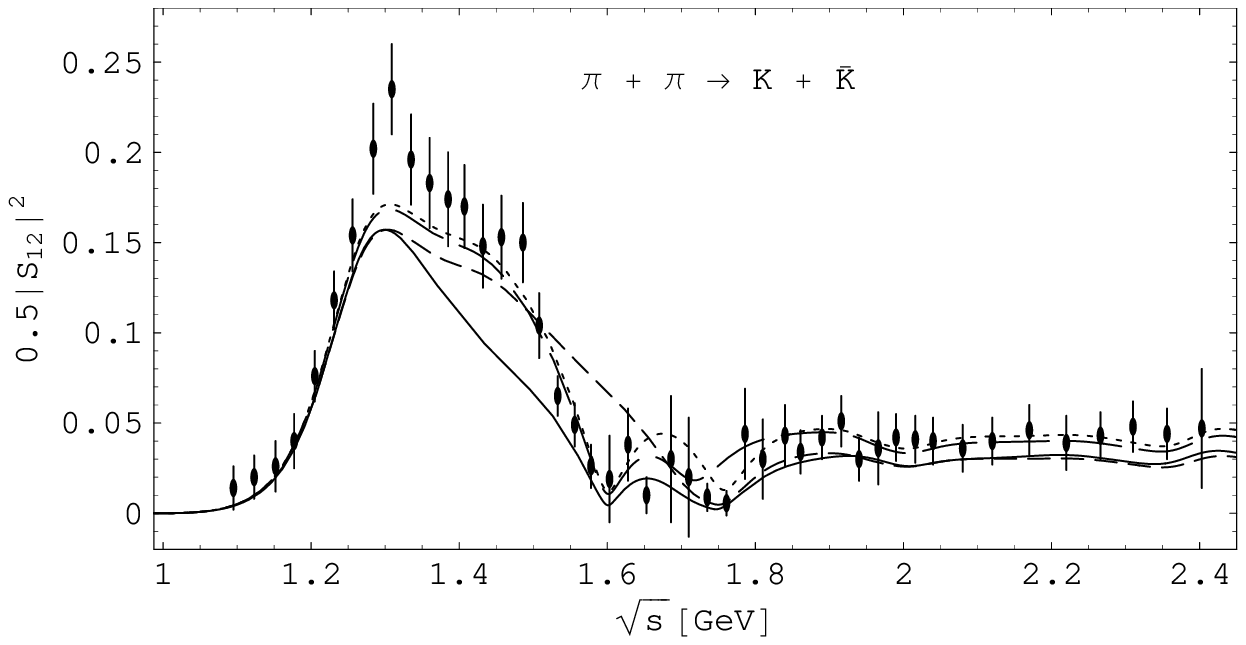}
\includegraphics[width=0.495\textwidth,angle=0]{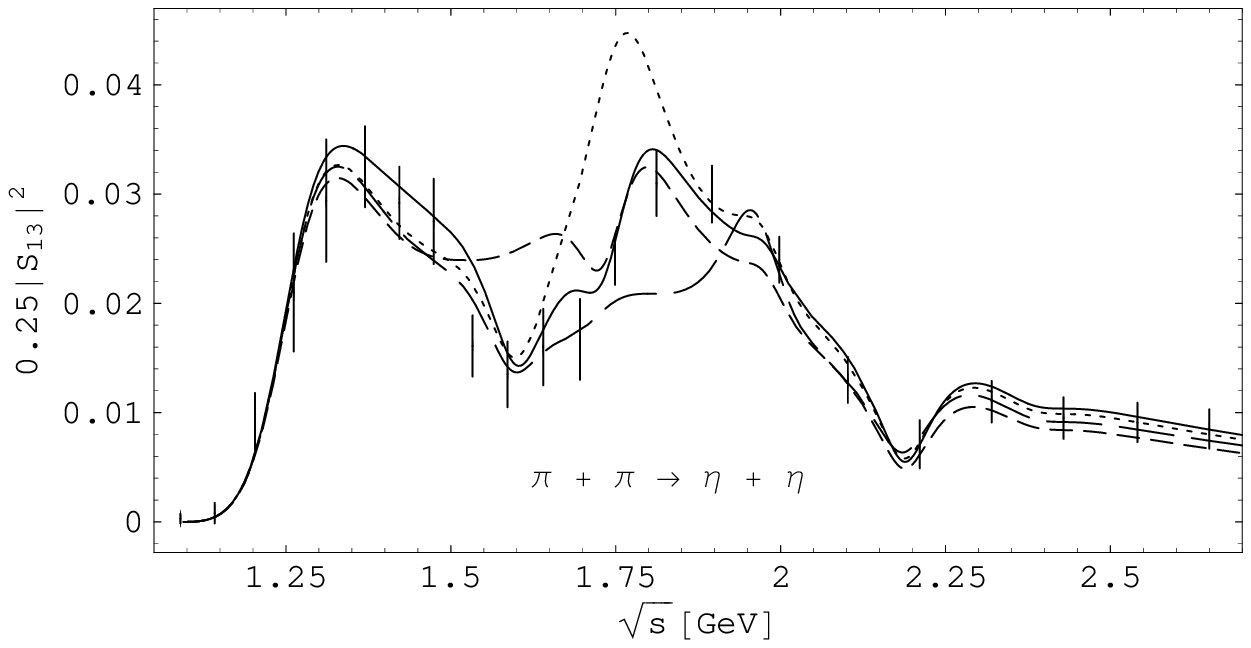}\\
\hspace*{1cm}(c) \hspace*{7.5cm} (d) \\
\caption{Scattering properties of the $\pi\pi \to \pi\pi, K\bar K, \eta\eta$ processes
for four scenarios:
(1) $f_2(1450.5)$ and $f_2(1534.7)$ states are excluded (solid line);
(2) $f_2(1534.7)$ and $f_2(1719.8)$ states are excluded (dotted line);
(3) $f_2(1534.7)$ and $f_2(1601.5)$ states are excluded (short-dashed line);
(4) $f_2(1534.7)$ and $f_2(1760)$ states are excluded (long-dashed line). \\
The experimental data are from Refs.~\cite{Hya73} and~\cite{Lin92}.\\
(a): Phase shift and in the $D$-wave $\pi\pi$-scattering. \\
(b): Modulus of the $S$-matrix element in the $D$-wave $\pi\pi$-scattering. \\
(c): Squared moduli of the $S$-matrix element in the $D$-wave $\pi\pi \to K \bar K$ process.\\
(d): Squared moduli of the $S$-matrix element in the $D$-wave $\pi\pi \to \eta\eta$ process.}
\label{24_23_25_27}
\end{center}
\end{figure}

\begin{figure}[!thb]
\begin{center}
\includegraphics[width=0.495\textwidth,angle=0]{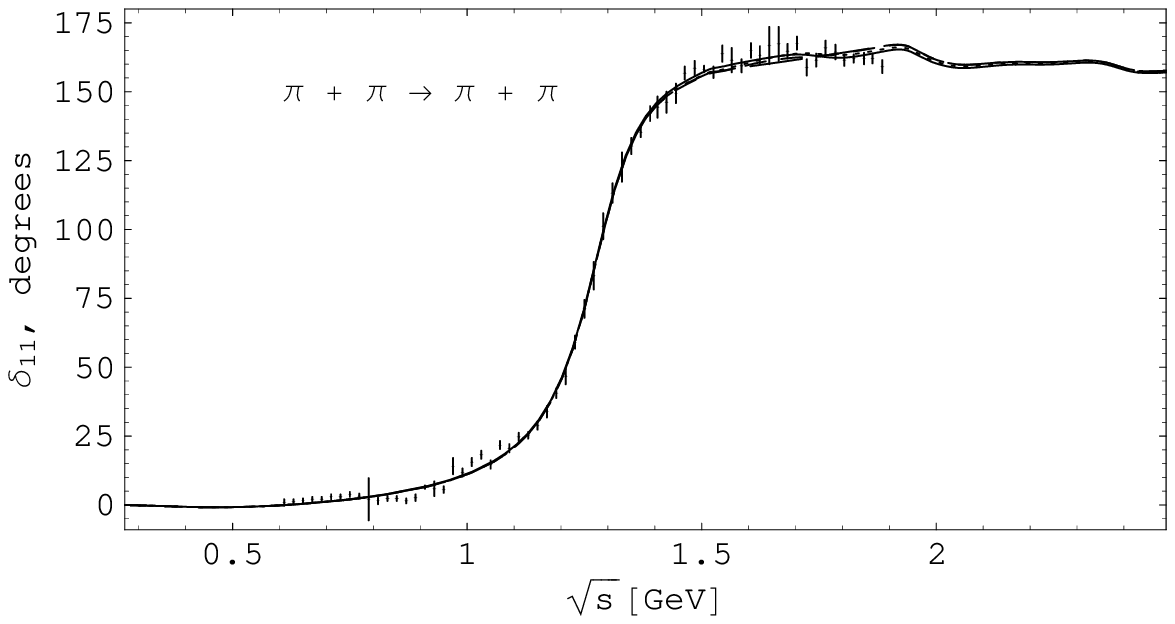}
\includegraphics[width=0.495\textwidth,angle=0]{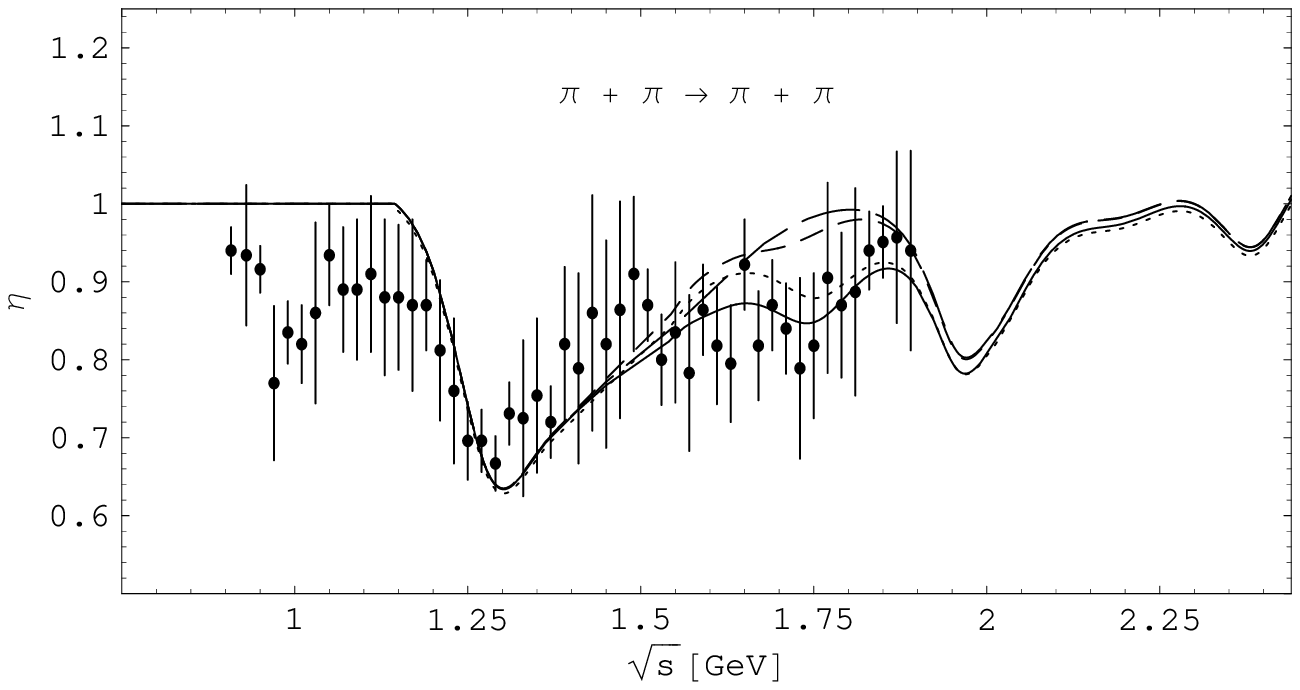}\\
\hspace*{1cm}(a) \hspace*{7.5cm} (b) \\[5mm]
\includegraphics[width=0.495\textwidth,angle=0]{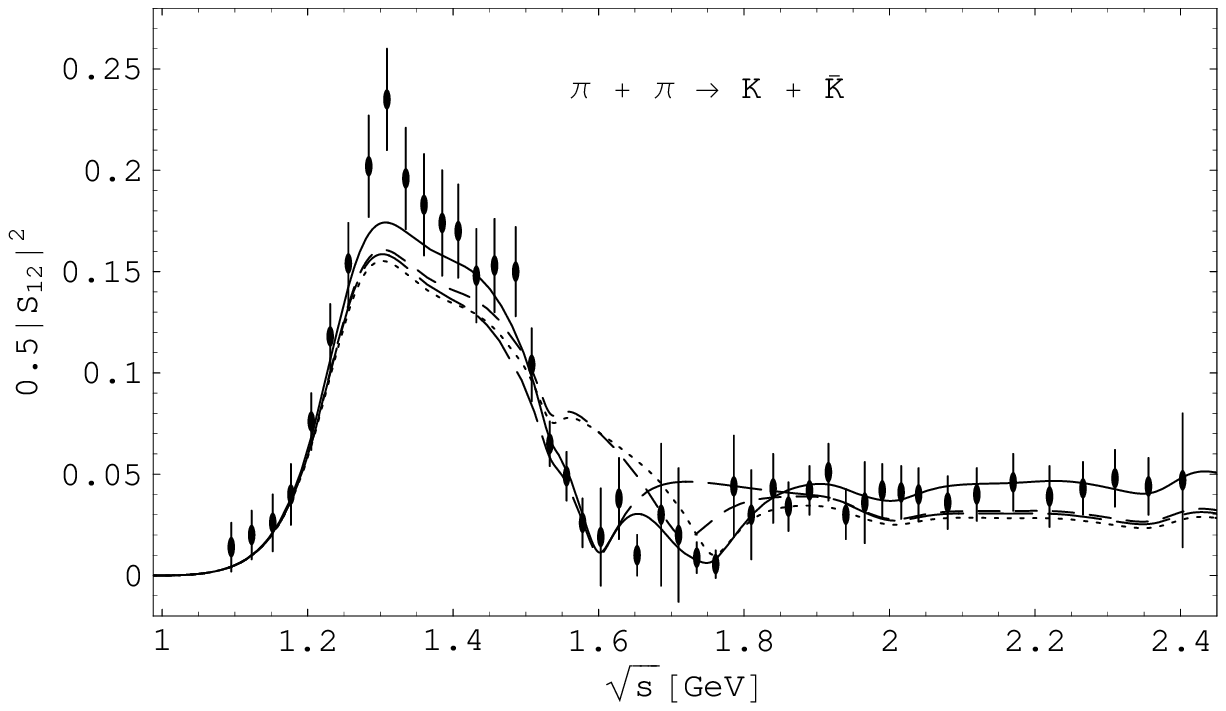}
\includegraphics[width=0.495\textwidth,angle=0]{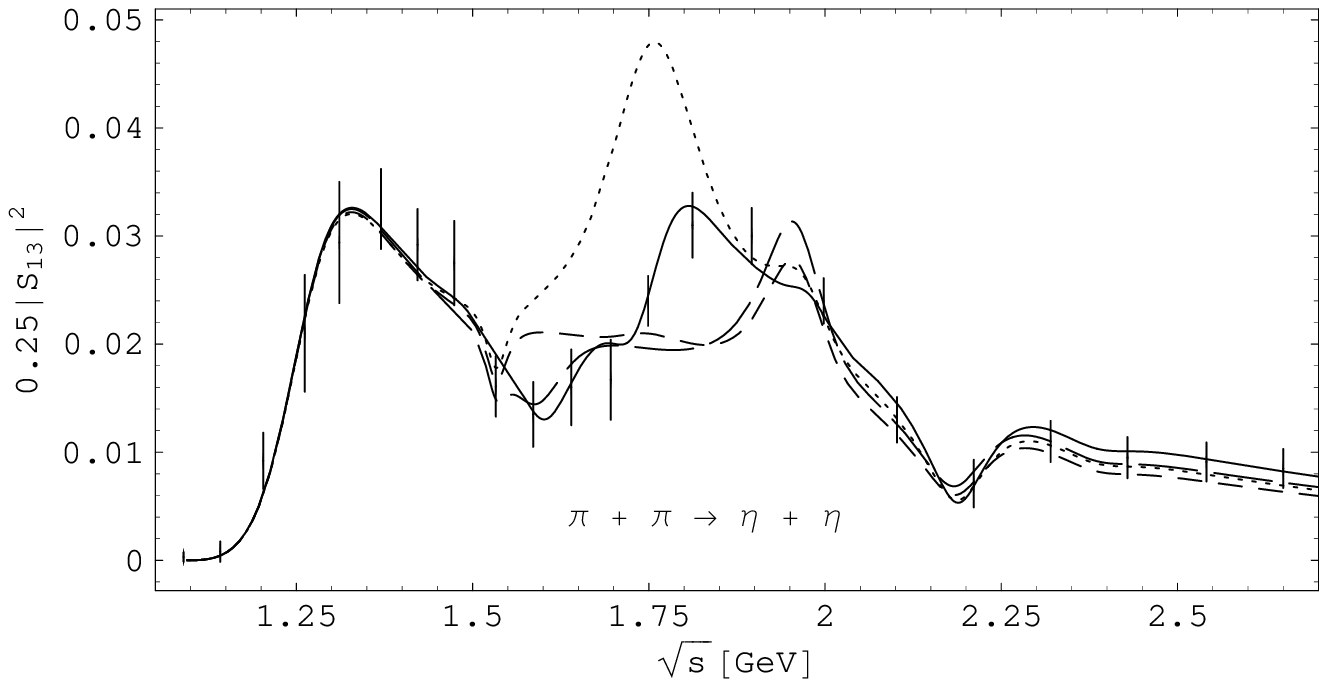}\\
\hspace*{1cm}(c) \hspace*{7.5cm} (d) \\
\caption{Scattering properties of the $\pi\pi \to \pi\pi, K\bar K, \eta\eta$ processes
for four scenarios:
(1) full result with inclusion of all relevant $f_2$-resonances (solid lines);
(2) $f_2(1601.5)$ and $f_2(1719.8)$ states are excluded (dotted line);
(3) $f_2(1601.5)$ and $f_2(1760)$ states are excluded (short-dashed line);
(4) $f_2(1760)$ and $f_2(1719.8)$ states are excluded (long-dashed line). \\
The experimental data are from Refs.~\cite{Hya73} and~\cite{Lin92}.\\
(a): Phase shift and in the $D$-wave $\pi\pi$-scattering. \\
(b): Modulus of the $S$-matrix element in the $D$-wave $\pi\pi$-scattering. \\
(c): Squared moduli of the $S$-matrix element in the $D$-wave $\pi\pi \to K \bar K$ process.\\
(d): Squared moduli of the $S$-matrix element in the $D$-wave $\pi\pi \to \eta\eta$ process.}
\label{all_35_37_57}
\end{center}
\end{figure}

Our original aim was to obtain a satisfactory combined description of the reaction $\gamma\gamma\to\pi^0\pi^0$,
of previously investigated decays of charmonia -- $J/\psi\to\phi(\pi\pi,K\overline{K})$,
$\psi(2S)\to J/\psi\,\pi\pi$, and $X(4260)\to J/\psi~\pi^+\pi^-$ -- of bottomonia --
$\Upsilon(mS)\to\Upsilon(nS)\pi\pi$ ($m>n$, $m=2,3,4,5,$ $n=1,2,3$)~\cite{SBGKLN-prd18}, and the
isoscalar $S$- and $D$-wave multi-channel $\pi\pi$-scattering processes~\cite{SBLKN-prd14}.
With respect to the last cases we also carried out calculations for the $D$-wave processes
$\pi\pi\to\pi\pi,(2\pi)(2\pi),K\overline{K},\eta\eta$ studying the various variants of
the $f_2$-resonances. Results are shown in Figures~\ref{all_2_4_7} -- \ref{all_35_37_57}. In these
figures it is seen that the phase shifts of $D$-wave $\pi\pi$-scattering are almost the same in all
the considered scenarios. All differences in the description of data are encoded
in the moduli of the amplitudes of studied processes.

\section{Conclusions}
We studied the process $\gamma\gamma\to\pi^0\pi^0$
and obtained a quite satisfactory description of the data~\cite{CrBall_gg,Belle13,Belle_gg} on the
cross section from the $\pi\pi$ threshold
up to $\approx2.5$~GeV with a $\chi^2/\mbox{n.p.}\approx 1.30$.

The energy spectrum of $\gamma\gamma\to\pi^0\pi^0$ from the $\pi\pi$-threshold
to the $K\overline{K}$-threshold is almost completely determined by the $S$-wave contribution and
above the $K\overline{K}$-threshold (and especially above $\approx1.3$~GeV) by the $D$-wave contribution.
Therefore, the observed bell-shaped behaviour of the energy spectrum in the near-$\pi\pi$-threshold
region is related to the $S$-wave contribution, whereas the structures above 1.3~GeV to
the $f_2$-resonances contributing to the $D$-wave including their interference (see Fig.~\ref{Partialcontr}(a).
From the calculated contributions of individual coupled channels to the energy spectrum of
$\gamma\gamma\to\pi^0\pi^0$ (Fig.~\ref{Partialcontr}(b)) we conclude that
the bell-shaped behaviour of the near-$\pi\pi$-threshold energy spectrum is mainly related to the
$S$-wave $K\overline{K}$ intermediate state contribution. There are also large contributions by the
$D$-wave process $\gamma\gamma\to\pi^+\pi^-\to\pi^0\pi^0$ and above $\approx0.85$~GeV
by the $D$-wave process $\gamma\gamma\to\eta\eta\to\pi^0\pi^0$. We also notice a sizable contribution
by one-pion exchange to the crossing channel of the $D$-wave process $\gamma\gamma\to\pi\pi\to\pi\pi$.
The intermediate state $\eta\eta$ in the reaction $\gamma\gamma\to\pi^0\pi^0$ implies, of course, a
preceding pair of charged particles ($\pi^+\pi^-$) in the intermediate state, that is, the process
$\gamma\gamma\to\pi^+\pi^-\to\eta\eta\to\pi^0\pi^0$ and the one-pion exchange between the $\gamma\pi^+$
and $\gamma\pi^-$ vertices. This explains the noticeable contribution of the
$D$-wave process $\gamma\gamma\to\eta\eta\to\pi^0\pi^0$.

For investigating the role of individual $f_2$-resonances in
$\gamma\gamma\to\pi^0\pi^0$  we switched off the different isoscalar-tensor
resonances and performed for each case the combined analysis of $\gamma\gamma\to\pi^0\pi^0$,
of the isoscalar $S$-wave processes $\pi\pi\to\pi\pi,K\overline{K},\eta\eta$ and of the isoscalar
$D$-wave processes $\pi\pi\to\pi\pi,(2\pi)(2\pi),K\overline{K},\eta\eta$. Results of the analyses
are shown in Table~\ref{tab:khi_sqrd} and Figures~\ref{all_2_4_7} -- \ref{all_35_37_57}.
The best two variants with the $f_2(1601.5)$ and $f_2(1719.8)$ and the $f_2(1601.5)$ and $f_2(1760)$,
switched off, do not give, however, a satisfactory description of the processes
$\pi\pi\to K\overline{K},\eta\eta$.

We finally completed (including previous work) a combined description of the isoscalar
$S$-wave processes $\pi\pi\to\pi\pi,K\overline{K},\eta\eta$, the isoscalar $D$-wave processes
$\pi\pi\to\pi\pi,(2\pi)(2\pi),K\overline{K},\eta\eta$, the decays of charmonia --
$J/\psi\to\phi(\pi\pi,K\overline{K})$, $\psi(2S)\to J/\psi\,\pi\pi$, and $X(4260)\to J/\psi~\pi^+\pi^-$,
of bottomonia -- $\Upsilon(mS)\to\Upsilon(nS)\pi\pi$ ($m>n$, $m=2,3,4,5,$ $n=1,2,3$),
and of the process $\gamma\gamma\to\pi^0\pi^0$.
In future we plan to include in our analysis data on the differential cross sections of
the $\gamma\gamma \to \pi^+ \pi^-$ process.

\vspace*{.5cm}
\funding{
This work was supported in part by the Heisenberg-Landau Program, by the Votruba-Blokhintsev Program
for Cooperation of Czech Republic with JINR, by the Czech Science Foundation GACR Grant No. 19-19640S (P.B.),
by the Grant Program of Plenipotentiary of Slovak Republic at JINR, by
the Bogoliubov-Infeld Program for Cooperation of Poland with JINR, by the BMBF (Project 05P2021,
F\"orderkennzeichen: 05P21VTCAA), by ANID PIA/APOYO AFB180002 (Chile),
by FONDECYT (Chile) under Grant No. 1191103, by ANID$-$Millennium Program$-$ICN2019\_044 (Chile),
and by the Polish research project with Project No. 2018/29/B/ST2/02576 (National Science Center).
}

\vspace*{.5cm}
\reftitle{References}

\end{document}